\documentclass[preprint,showpacs,preprintnumbers,amsmath,amssymb]{revtex4}

\usepackage{graphicx}
\usepackage{dcolumn}
\usepackage{bm}

\begin{document}

\title{Nonstationary theory of magnetic field induced current for molecular spin nanojunction}

\author{M. Jouravlev$^{1}$ and Kwang S. Kim$^{2}$}

\affiliation{Center for Superfunctional Materials, Department of Chemistry, Pohang University of Science and Technology,
Pohang 790-784, Korea}
\email{1) mikhail@postech.ac.kr, 2) kim@postech.ac.kr}

\begin{abstract}
For the study of molecular spin junctions, we take into account two types of couplings between the
molecule and the metal leads: (i) electron transfer that gives rise to net current
in the biased junction and (ii) energy transfer between the molecule and the leads. Using a rotating wave
approximation in the Heisenberg representation, we
derive a set of differential equations for the expectation values of relevant
variables: electron and phonon populations and molecular polarization.
A magnetic field control method to enhance the charge transfer at spin nanojunctions,
which characterizes the molecule feature, is discussed. An approximate analytical solution of the resulting dynamical equation is
supported by numerical solution. The magnetic control by charge transfer is
described by transient pseudo-fermions of electrons interacting with spins.
The rapid adiabatic passage of the energy between the molecule and the leads is taken into account.
The current for molecular spin nanojunctions is derived.
\end{abstract}

\maketitle

\section{Introduction}
\label{sec:Introduction}
A molecular spin nanojunction is a nanodevice where
the electron transfer depends on the spin state of electrons passing through the
molecule and is controlled by an external magnetic field.
The giant magnetoresistance\cite{Binasch,Baibich,Parkin,Asano} and tunneling magnetoresistance\cite{Moodera1,Moodera2,Mehrez}
are widely used as molecular memory devices for magnetic field recording, and this has
launched a new field of nanoelectronics - "spintronics".
Recently, even super magnetoresistance has been discussed through graphene nanoribbons\cite{KimWY}.

The main theoretical principle to describe the properties
of tunneling electron transfer was first proposed by Gamov\cite{Gam}.
In the last few years some theoretical and computational aspects of electron conduction in nanojunctions
have been under intense study\cite{Kagan_Ratner04,Joachim_Ratner05,Hanggi02,Hanggi04,Kim3}.
The electric-field driven magnetic switching has been discussed\cite{Dief} and
the light induced switching behavior in the conduction properties of molecular
nanojunctions has been demonstrated \cite{Dulic03,Katsonis06,He05,Wakayama04,Yasutomi04}.
The formalism of the quantized electron conductance was derived
\cite{Land1,Landauer1} and the conduction behavior in the heterostructure of molecular nanojunction was formulated.
The spin nanojunction for carbon nanotubes or graphene nanoribbons was investigated\cite{Kim2,Kim5}.
Spin dependent electron transport was seen in electron nanotubes and graphene excited by nonstationary
magnetic field\cite{Sungjae} as well as in graphene nanoribbons doped by
chemically active impurities\cite{Martins,Pisani}.

The formalism of the current in molecular nanojunctions was derived based on elastic electron scattering between two electronic
baths corresponding to two leads\cite{Nit05,Nitzan06JCP,Nit7}.
In the elastic scattering limit, energy is lost in the lower potential lead, while no energy dissipation occurs
in the molecular nanojunction. Following the Landauer formalism\cite{Land1,Landauer1,Butt,Butt2},
 most of theoretical works on the nanojunction
transport were done within a scattering theory approach,
which neglects the contact problems and the influence of
the scattering channels as well as the mutual influences between
the electron and phonon subsystems\cite{Naa,Brand,Larad}.

The electron transfer rate in terms of the coupling between the electronic state and the nuclear vibration was provided\cite{Jort}.
The spatial resolution at the atomic scale for single molecules adsorbed on the surface
has also been achieved by scanning tunneling microscopy (STM)\cite{NitGal}.
The nuclear vibration is described by transversal time or contact time for the electron
transfer through a molecule. In order to estimate the transversal time, it is necessary to take into account the chain length
of a molecule in a small gap between two leads. The transfer time can be long enough to be comparable to the molecular vibration period\cite{NitJor}.
The strong coupling between molecular vibration and electronic states could result in trapping of electrons and
transition from the coherent to incoherent state\cite{Nit10}. This transition has been found in
molecular nanowires under the radiation field. It could be achieved by both long range electron transfers and currents through the molecular nanojunctions in nonequilibrium states\cite{Caron}. A class of molecules characterized by strong charge-transfer driven
transitions into their first excited state has been investigated\cite{Nit05}.
The dipole moment of such molecules changes considerably upon the excitation,
resulting in a substantial electronic charge redistribution.

Here, we focus our attention on the transport of the electron through the single-molecule spin nanojunction
under the external magnetic field which controls a small nanogap between two metal leads.
This research effort is now devoted to extending the spin-dependent effects to magnetic molecular nanojunction
for spintronic nanodevices with relatively strong
electron-phonon coupling and large spin coherence.
We investigate the spin polarized electron transport in either occupied (OMO) or unoccupied molecular orbitals (UMO),
(Fig.\ref{fig:model}). For a molecular spin nanojunction connecting between two
metal leads, time dependent magnetic fields can create an internal driving
force for the charge to flow between the two leads. We suppose that the molecular junction has extremely weak spin-orbit interaction
and weak hyper-fine interaction, meaning that the electron spin diffusion length is long enough to provide
the spin-polarized electron injection and the spin transport between the leads.
Our objective in the present work is to extend the theory of Galperin, Fainberg and Nitzan\cite{Nit05,Nitzan06JCP,Fain}
to the case including magnetic fields and to apply the theory to the study of coherent control of nanojunction transport.
While these problems are of general and fundamental interest, we note that
this study is related to the efforts to develop novel single-electron devices, magnetic memory devices, and single-electron transistors with magnetic gating\cite{Wakayama04}. In addition, the potential significance of molecular spin nanojunctions for device
applications lies in the possibility of creating magnetic switches
 that could be incorporated in future generations of
communication systems\cite{Hanggi03}. It is conceivable that these devices will employ
coherent spin manipulations for quantum information processing.

In this study we particularly develop theory for the effects of strong electron-phonon interactions
on tunnelling nanojunction and inelastic tunnelling in quantum
point contacts associated with the non-linear conduction phenomena.
There are reasons to consider molecular nanojunctions subjected to strong
magnetic fields. First, the structure of such junctions is
compatible with the configuration considered for high electromagnetic
field as in tip enhanced scanning near field optical microscopy (SNOM)\cite{Keilmann}.
Second, consideration of the spin nanojunction stability suggests that strong radiation
fields should be applied in the sequence of well separated pulses to
allow for sufficient relaxation and heat dissipation. Finally,
consideration of strong time dependent pulses makes it possible to
study the way to optimize the desired effect for the magnetic field induced
electron tunneling, i.e. to explore the possibility for coherent
control of charge flow between the leads.

This paper is organized as follows. Sec.\ref{sec:Hamiltonian} introduces our model.
Sec.\ref{sec:equations} derives a closed set of
Heisenberg equations for the average values for the magnetic spin operators and for
the annihilation and creation operators for the electrons in the molecular states.
In Sec. \ref{sec:current} and \ref{sec:diffeq}, we derive the current functions for
the magnetic spin operators and the formulas for the current and charge
transfer during the magnetic pulse action, and
calculate the current induced by a quasi-stationary
magnetic pulse. The results for control of the current and the transferred charge by
chirped pulses are summarized in Sec.\ref{sec:Magnetcont}, and the conclusion is made in Sec.\ref{sec:Concl}.
In Appendices, we show that in the absence of the radiative and nonradiative energy transfer couplings, the
equations of motion derived here lead to the well known Landauer's type
formula for the current.

\section{Hamiltonian}
\label{sec:Hamiltonian}
\begin{figure}[ptb]
\begin{center}
\includegraphics[width=3.4in]%
{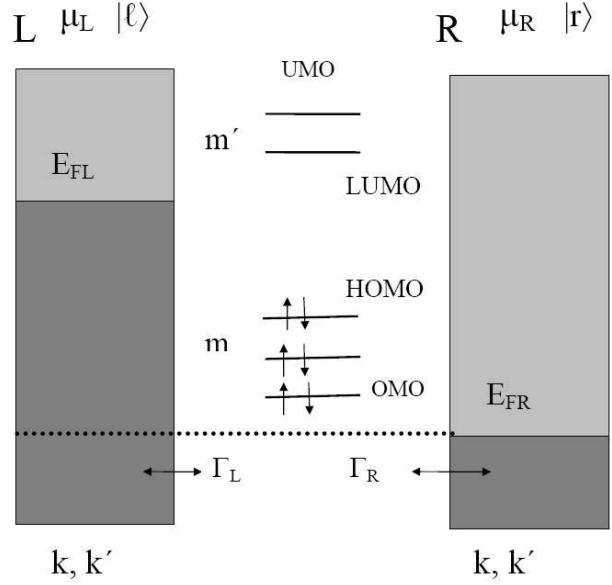}%
\caption{A model for the molecule-electrode junction. The electronic structure is
substantially modified as the molecule is between two leads by a nanojunction.
The right ($R=|\{r\}\rangle$ ) and left ($L=|\{l\}\rangle$) manifolds represent two
metal leads characterized by electrochemical potentials $\mu_{R}$ and $\mu_{L}$,
respectively. Occupied
molecular orbitals (OMO) and unoccupied molecular orbitals
(UMO) are schematically drawn. The Fermi energy of the left electrode $(E_{FL})$ is higher
than the Fermi energy of the right electrode $(E_{FR})$ in the presence of magnetic field.
In the Hamiltonians Eqs.(\ref{eq:Ho}, \ref{eq:V_M}, and \ref{eq:V_N}), indices $k,k'$ denote the states of the metal leads,
and indices $m,m'$ denote the states of the molecule. Couplings $\Gamma_{L}$ and $\Gamma_{R}$ are denoted by double
headed arrows.}%
\label{fig:model}%
\end{center}
\end{figure}

As for the simplest theoretical view of an efficient spin molecular transport system, the actual contacts
of the molecule to both electrodes are presented in Fig.\ref{fig:model}.
The metal electrodes are on the left ($L$) and right ($R$) sides.
The occupied molecular orbitals (OMO) and unoccupied molecular orbitals (UMO) are presented by
line segments, and the spins are depicted by arrows.
A molecule is positioned between two leads represented by free electron
reservoirs $L$ and $R$. The system interacts with the magnetic field. The reservoirs are
characterized by the electronic chemical potentials $\mu_{L}$ and $\mu_{R}$,
where the difference $\mu_{L}-\mu_{R}$ $=eV$ is the imposed voltage bias.
The coupling $\Gamma$ is shown by double headed arrows.
In the independent electron picture, the transition
between the ground and excited molecular states corresponds to the transfer of an
electron between OMO and UMO levels. There are couplings between atomic levels
and electrodes. The Fermi level of the electrodes lies within the gap between the highest OMO
(HOMO) and the lowest UMO (LUMO) of the molecule. This picture assumes that the coupling between
a molecule and an electrode is relatively weak compared with the interatomic interactions.
The magnetization direction is either parallel or perpendicular to
 the external magnetic field $B(r,t)$.
In Fig.\ref{fig:model}, the model Hamiltonian Eq.(\ref{eq:H}) consists of the unperturbed part
$(\hat{H}_{0})$, the perturbed part $(\hat{V})$, and the magnetic field effect part $(\hat{H}_{B})$:
\begin{equation}
\hat{H}=\hat{H}_{0}+\hat{H}_{B}+\hat{V}, \label{eq:H}%
\end{equation}
where
\begin{eqnarray}
\hat{H}_{0}=\sum_{\alpha}\omega_{\alpha}\hat{a}_{\alpha}^{+}\hat{a}_{\alpha}+
\sum_{\beta}\omega_{\beta}\hat{b}_{\beta}^{+}\hat{b}_{\beta}\nonumber\\
+\sum_{m\in M,\sigma}\varepsilon_{m}\hat{c}_{m,\sigma}^{+}\hat{c}_{m,\sigma}+\sum
_{k\in\{L,R\},\sigma}\varepsilon_{k}\hat{c}_{k,\sigma}^{+}\hat{c}_{k,\sigma}.
\label{eq:Ho}%
\end{eqnarray}
Here, $L$ and $R$ denote the left and right leads, respectively.
We use $\hbar=1$ and $e=1$ here and below. The Hamiltonian $\hat{H}_{0}$ (Eq.(\ref{eq:Ho})) contains additive terms that correspond to the isolated molecule (index: m) and the free leads (index: k). Operators $\hat{c}_{m,\sigma}$ ($\hat{c}^{+}_{m,\sigma}$) and $\hat{c}_{k,\sigma}$ ($\hat{c}^{+}_{k,\sigma}$) are an annihilation (creation) operator of an electron in the molecule and that in the leads, respectively,
where $\sigma$ is the spin ($\uparrow$ or $\downarrow$); operators $\hat{a}_{\alpha}$
($\hat{a}^{+}_{\alpha}$) and $\hat{b}_{\beta}$ ($\hat{b}^{+}_{\beta}$) are an annihilation (creation) operator of a phonon
in the molecule and that in the thermal bath or phonon reservoir, respectively.
The interaction part $\hat{V}$ of Hamiltonian (Eq.(\ref{eq:H})) is:
\begin{equation}
\hat{V}=\hat{V}_{M}+\hat{V}_{N}+\hat{V}_{V}.\label{eq:V}%
\end{equation}
Here the $\hat{V}_{M}$ term describes the coupling between the molecular electronic
subsystem and the free-electron reservoirs in the leads, and $\hat{V}_{M}$ has the form\cite{MeirAnd}:
\begin{equation}
\hat{V}_{M}=\sum_{K=L,R}\sum_{m\in M;k\in K,\sigma}(V_{km}^{(MK)}\hat{c}_{k,\sigma}^{+}%
\hat{c}_{m,\sigma}+h.c.).\label{eq:V_M}%
\end{equation}
Here h.c. denotes Hermitian conjugate. The term $\hat{V}_{N}$ of the Hamiltonian (Eq.\ref{eq:V}) describes the energy transfer between the molecule and the leads. It is written in the form of Anderson Hamiltonian\cite{Ander,Mah,MeirAnd}:
\begin{equation}
\hat{V}_{N}=\sum_{K=L,R}\sum_{m\neq m^{\prime}\in M;k\neq k^{\prime}\in K,\sigma}(V_{kk^{\prime}}
^{(NK)}\hat{c}_{k,\sigma}^{+}\hat{c}_{k^{\prime},\sigma}\hat{c}_{m,\sigma}^{+}\hat{c}_{m^{\prime},\sigma}).
\label{eq:V_N}
\end{equation}
Electrons of the nanojunction are coupled to both the vibrations
of the molecule and the electrons of the leads. The $\hat{V}_{V}$ term is the coupling
potential of the interaction between electrons in the molecule and phonons in the subsystem,
which is taken to be linear to the vibrational displacements in the
form of Fr\"{o}hlich Hamiltonian\cite{Mah}:
\begin{equation}
\hat{V}_{V}=\sum_{m\neq m^{\prime}\in M,\sigma}(V_{mm^{\prime}}
^{(VK)}\hat{c}_{m,\sigma}^{+}\hat{Q}_{\alpha}^{a}\hat{c}_{m^{\prime},\sigma}+h.c.).
\end{equation}
Here, $\hat{Q}_{\alpha}^{a}$ is the vibrational displacement operator:
$\hat{Q}_{\alpha}^{a}=\hat{a}_{\alpha}+\hat{a}_{\alpha}^{+}$.
Magnetic interactions in the Heisenberg model are employed as effective interactions between the spin of the electron
and the magnetic fields. The model of spin Hamiltonian is described by magnetic interactions with a few model parameters such as
 spin-spin interaction\cite{Kad}:
\begin{equation}
\hat{H}_{S}=-2\sum_{m>m',\sigma\sigma'}J_{m,m'}S_{m}S_{m'},
\label{eq:Hs}
\end{equation}
which is called Ising spin Hamiltonian\cite{Kad,Kad2}. Here, $S_{m}$ is the localized spin magnetic moment, $J_{m,m'}$
is the interaction potential for which each pair of spins $S_{m}$ and $S_{m'}$ is counted only once. $J_{m,m'}$ is the form of
Green functions consistent with the spin fluctuation theory based on Kohn-Sham eigenfunctions\cite{Gop,Hirosha}.
However it is dropped out because it is a constant independent of the magnetic field.
The Hamiltonian has the form of Zeeman term\cite{Nit1} with the
external magnetic field $B(r,t)$ for the chosen value of the g-factor
with the pseudo-Fermion operators\cite{Wolf,Grab,Gop}:
\begin{equation}
\hat{H}_{B}=\mathbf{H}_{B}(\mathbf{r},t)\hat{c}_{m,\sigma}^{+}\hat{c}_{m',\sigma'},
\label{eq:hb1}
\end{equation}
where
\begin{equation}
\mathbf{H}_{B}(\mathbf{r},t)=g\mu_{B}\sum_{m\in M,\sigma}\mathbf{B}(\mathbf{r},t)S_{m}.
\label{eq:hb2}
\end{equation}
Here, $g$ is the Lande g-factor which is normally close to 2;
$\mu_{B}$ is the Bohr magneton  and $B(\mathbf{r},t)$ is the magnetic field. The $\hat{c}_{m,\sigma}^{+}$ and $\hat{c}_{m,\sigma}$
operators  describe pseudo-Fermion operators specified by $\sigma$ which raise the energies by $\pm g\mu_{B}B(\mathbf{r},t)$.
Taking into account that electrons with the pseudo-Fermion properties exist only in the presence  of
the magnetic fields, we omit the index $\sigma$ in all formulas below.

\section{Heisenberg equations}
\label{sec:equations}

The physics of the system can be described by different approaches. One is
the method of nonequilibrium Green's functions
\cite{Nit05,Nitzan06JCP,Muk2}. It has advantages in formal
treatment due to a diagrammatic representation, and it is
particularly well suited for stationary processes where the Dyson equation can
be cast in the energy representation. For time-dependent processes, a method based on the Heisenberg equations of motion for
the expectation values of the operators provides a more transparent approach,
since the quantities are more directly related to physical observables. Such a
method is adopted here. The Heisenberg equations for $\hat{c}_{m}$$(\hat{c}_{m}^{+})$
and $\hat{c}_{k}$$(\hat{c}_{k}^{+})$
can be written as follows:
\begin{equation}
\frac{d\hat{c}_{m}}{dt}=\frac{i}{\hbar}[\hat{H}_{0}+\hat{H}_{B}+\hat{V}_{M}+%
\hat{V}_{N}+\hat{V}_{V},\hat{c}_{m}], \label{eq:Heis1}%
\end{equation}
i.e.,
\begin{eqnarray}
\frac{d\hat{c}_{m}}{dt}=\frac{i}{\hbar}[\hat{H}_{0},\hat{c}_{m}]
+\frac{i}{\hbar}[\hat{H}_{B},\hat{c}_{m}]\nonumber\\
+\frac{i}{\hbar}[\hat{V}_{M},\hat{c}_{m}]+\frac{i}{\hbar}[\hat{V}_{N},\hat{c}_{m}]
+\frac{i}{\hbar}[\hat{V}_{V},\hat{c}_{m}].
\label{eq:Heis2}
\end{eqnarray}
The equation of motion derived by the mathematical transformation is presented in Appendices \ref{sec:AppA}-
\ref{sec:AppC}:
\begin{widetext}
\begin{eqnarray}
\frac{d\hat{c}_{m}}{dt} = -\frac{i}{\hbar}\varepsilon_{m}\hat{c}_{m}%
 -\frac{i}{\hbar}\sum_{K=L,R}\sum_{k\in K}V_{mk}^{(MK)}\hat{c}_{k}%
 -\frac{i}{\hbar}\sum_{K=L,R}\sum_{k\neq k^{\prime}\in K}\{V_{kk^{\prime}%
}^{(NK)}\hat{c}_{k}^{+}\hat{c}_{k^{\prime}}\hat{c}_{1}\delta_{2m}%
+V_{k^{\prime}k}^{(NK)}\delta_{1m}\hat{c}_{2}\hat{c}_{k^{\prime}}^{+}\hat%
{c}_{k}\}\nonumber\\
-\frac{i}{\hbar}\sum_{m\neq m^{\prime}\in M}(V_{mm^{\prime}}%
^{(VK)}\hat{Q}_{\alpha}^{a})\hat{c}_{m}%
+\frac{i}{\hbar}\mathbf{H}_{B}(\mathbf{r},t)\hat{c}_{m'\neq m}\text{,}%
\label{eq:cm}
\end{eqnarray}
\end{widetext}
and
\begin{widetext}
\begin{eqnarray}
\frac{d\hat{c}_{m}^{+}}{dt}=\frac{i}{\hbar}\varepsilon_{m}\hat{c}_{m}
^{+}+\frac{i}{\hbar}\sum_{K=L,R}\sum_{k\in K}V_{km}^{(MK)}\hat{c}_{k}
^{+}+\frac{i}{\hbar}\sum_{K=L,R}\sum_{k\neq k^{\prime}\in K}\{V_{k^{\prime}
k}^{(NK)}\hat{c}_{1}^{+}\hat{c}_{k^{\prime}}^{+}\hat{c}_{k}\delta
_{2m}+V_{kk^{\prime}}^{(NK)}\delta_{1m}\hat{c}_{k}^{+}\hat{c}_{k^{\prime}}
\hat{c}_{2}^{+}\}\nonumber\\
-\frac{i}{\hbar}\sum_{m\neq m^{\prime}\in M}(V_{mm^{\prime}}
^{(VK)}\hat{Q}_{\alpha}^{a})\hat{c}_{m}^{+}
-\frac{i}{\hbar}\mathbf{H}_{B}^{\ast}(\mathbf{r},t)\hat{c}_{m'\neq m}^{+}.
\label{eq:cpm}
\end{eqnarray}
\end{widetext}

\subsection{Calculation of electron transfer without external field}
\label{sec:without_field1}
We calculate the second term on the right-hand side of Eqs. (\ref{eq:cm}) and (\ref{eq:cpm}).
To do this, we shall write down the Heisenberg equation for $\hat
{c}_{k}$, when  $\hat{V}_{N}$ and $\hat{V}_{V}$ terms are not included:
\begin{equation}
\frac{d\hat{c}_{k}}{dt}=\frac{i}{\hbar}[\hat{H}_{0},\hat{c}_{k}]+\frac
{i}{\hbar}[\hat{V}_{M},\hat{c}_{k}]
\end{equation}
or
\begin{equation}
\frac{d\hat{c}_{k}}{dt}=-\frac{i}{\hbar}\varepsilon_{k}\hat{c}_{k}-\frac
{i}{\hbar}\sum_{m^{\prime}\in M}V_{km^{\prime}}^{(MK)}\hat{c}_{m^{\prime}}
\end{equation}
and
\begin{equation}
\frac{d\hat{c}_{m}}{dt}=-\frac{i}{\hbar}\varepsilon_{m}\hat{c}_{m}-\frac
{i}{\hbar}\sum_{k^{\prime}\in K}V_{k^{\prime}m}^{(MK)}\hat{c}_{k^{\prime}}.
\end{equation}
Since $\frac{d}{dt}(\hat{c}_{k}^{+}\hat{c}_{m})=\frac{d\hat{c}_{k}^{+}}%
{dt}\hat{c}_{m}+\hat{c}_{k}^{+}\frac{d\hat{c}_{m}}{dt}$, we obtain%
\begin{equation}
\frac{d\hat{c}_{k}^{+}}{dt}\hat{c}_{m}=\frac{i}{\hbar}\varepsilon_{k}\hat
{c}_{k}^{+}\hat{c}_{m}+\frac{i}{\hbar}\sum_{m^{\prime}\in M}V_{m^{\prime}%
k}^{(MK)}\hat{c}_{m^{\prime}}^{+}\hat{c}_{m}.%
\end{equation}
and%
\begin{equation}
\hat{c}_{k}^{+}\frac{d\hat{c}_{m}}{dt}=-\frac{i}{\hbar}\varepsilon_{m}\hat
{c}_{k}^{+}\hat{c}_{m}-\frac{i}{\hbar}\sum_{K=L,R}\sum_{k^{\prime}\in
K}V_{mk^{\prime}}^{(MK)}\hat{c}_{k}^{+}\hat{c}_{k^{\prime}}.%
\end{equation}
Then
\begin{eqnarray}
\frac{d}{dt}\langle\hat{c}_{k}^{+}\hat{c}_{m}\rangle=  %
\frac{i}{\hbar}(\varepsilon_{k}-\varepsilon_{m})\langle\hat{c}_{k}^{+}\hat{c}_{m}%
\rangle\nonumber\\
+\frac{i}{\hbar}\sum_{m^{\prime}\in M}V_{m^{\prime}k}^{(MK)}\langle%
\hat{c}_{m^{\prime}}^{+}\hat{c}_{m}\rangle\nonumber\\%
-\frac{i}{\hbar}\sum_{K=L,R}\sum_{k^{\prime}\in K}V_{mk^{\prime}}^{(MK)}\langle\hat{c}_{k}^{+}\hat%
{c}_{k^{\prime}}\rangle\label{eq:c^+_kc_m}%
\end{eqnarray}
where $\langle...\rangle$ denotes the averaging. Here we used $\langle\hat{c}_{m^{\prime}}^{+}\hat{c}_{m}\rangle=\langle\hat
{c}_{m}^{+}\hat{c}_{m}\rangle\delta_{mm^{\prime}}$ in the second term on the
right-hand side of Eq.(\ref{eq:c^+_kc_m}) because of the following reasons. First,
Eq.(\ref{eq:c^+_kc_m}) was obtained by neglecting $\hat{V}_{N}$ when the
polarization $\langle\hat{c}_{m^{\prime}}^{+}\hat{c}_{m}\rangle|_{m\neq
m^{\prime}}=0$. Second, the term $\langle\hat{c}_{m^{\prime}%
}^{+}\hat{c}_{m}\rangle|_{m\neq m^{\prime}}$ is the non-resonant case.
As to $\langle\hat{c}_{k}^{+}\hat{c}_{k^{\prime}}\rangle$, this term is equal to
\begin{equation}
\langle\hat{c}_{k}^{+}\hat{c}_{k^{\prime}}\rangle=f_{K}(\varepsilon_{k}%
)\delta_{kk^{\prime}},\label{eq:c^+_kc_k'}
\end{equation}
where $f_{K}(E)=[\exp((E-\mu_{K})/k_{B}T)+1]^{-1}$ is the Fermi function, $E$ is the energy, $k_{B}$ is the Boltzmann constant,
and $T$ is the temperature. Using $\langle\hat{c}_{m^{\prime}}^{+}\hat{c}_{m}\rangle=\langle\hat
{c}_{m}^{+}\hat{c}_{m}\rangle\delta_{mm^{\prime}}\equiv n_{m}$ and Eq.(\ref{eq:c^+_kc_k'})
for Eq.(\ref{eq:c^+_kc_m}), we obtain
\begin{eqnarray}
\frac{d}{dt}\langle\hat{c}_{k}^{+}\hat{c}_{m}\rangle=\frac{i}{\hbar
}(\varepsilon_{k}-\varepsilon_{m})\langle\hat{c}_{k}^{+}\hat{c}_{m}
\rangle\nonumber\\
+\frac{i}{\hbar}V_{mk}^{(MK)}[n_{m}-f_{K}(\varepsilon_{k})].
\label{eq:dc^+_kc_m}%
\end{eqnarray}
By integrating Eq.(\ref{eq:dc^+_kc_m}), we obtain%
\begin{widetext}
\begin{equation}
\langle\hat{c}_{k}^{+}\hat{c}_{m}\rangle=\frac{i}{\hbar}V_{mk}^{(MK)}%
\int_{-\infty}^{t}dt^{\prime}\exp[\frac{i}{\hbar}(\varepsilon_{k}%
-\varepsilon_{m})(t-t^{\prime})][n_{m}(t^{\prime})-f_{K}(\varepsilon_{k})].
\end{equation}
\end{widetext}
By using the formula
\begin{equation}
\int_{0}^{\infty}d\tau\exp[\frac{i}{\hbar}(\varepsilon_{k}-\varepsilon
_{m})\tau]=\frac{i\hbar P}{\varepsilon_{k}-\varepsilon_{m}}+\hbar\pi
\delta(\varepsilon_{k}-\varepsilon_{m}),
\end{equation}
where $P$ denotes the principal value and by assuming that the term $[n_{m}(t^{\prime})-f_{K}(\varepsilon_{k})]$ is slowly
varying as compared to the exponential function, we can move the term outside the integral of (Eq.\ref{eq:dc^+_kc_m}).
The resulting integral gives
\begin{widetext}
\begin{eqnarray}
\langle\hat{c}_{k}^{+}\hat{c}_{m}\rangle\simeq\frac{i}{\hbar}V_{mk}%
^{(MK)}[n_{m}(t)-f_{K}(\varepsilon_{k})]\int_{0}^{\infty}d\tau\exp[\frac
{i}{\hbar}(\varepsilon_{k}-\varepsilon_{m})\tau]\nonumber\\
=\frac{i}{\hbar}V_{mk}^{(MK)}[n_{m}(t)-f_{K}(\varepsilon_{k})][\frac{i\hbar
P}{\varepsilon_{k}-\varepsilon_{m}}+\hbar\pi\delta(\varepsilon_{k}%
-\varepsilon_{m})].\label{eq:c^+_kc_m_1}%
\end{eqnarray}
\end{widetext}
Let us show the important result of the simplified equations for the current $I$.
 Taking into account only $\hat{V}_{M}$ ($V_{mk}^{(MK)}$) in Eq.(\ref{eq:c^+_kc_m_1}), and
substituting the last result into $\frac{2}{\hbar}\operatorname{Im}%
\sum_{K=L,R}\sum_{k\in K}V_{km}^{(MK)}\langle\hat{c}_{k}^{+}\hat{c}_{m}%
\rangle$, we have
\begin{widetext}
\begin{align}
\frac{2}{\hbar}\operatorname{Im}\sum_{K=L,R}\sum_{k\in K}V_{km}^{(MK)}%
\langle\hat{c}_{k}^{+}\hat{c}_{m}\rangle &  =\frac{2\pi}{\hbar}\sum
_{K=L,R}\sum_{k\in K}|V_{km}^{(MK)}|^{2}[n_{m}(t)-f_{K}(\varepsilon
_{k})]\delta(\varepsilon_{k}-\varepsilon_{m})\nonumber\\
&  =\sum_{K=L,R}[n_{m}(t)-f_{K}(\varepsilon_{m})]\Gamma_{MK,m}.
\label{eq:Term_current}%
\end{align}
\end{widetext}
where%
\begin{equation}
\Gamma_{MK,m}=\frac{2\pi}{\hbar}\sum_{k\in K}|V_{km}^{(MK)}|^{2}%
\delta(\varepsilon_{k}-\varepsilon_{m}). \label{eq:GammaMKm}%
\end{equation}
Here we use the formula for current $I$, derived from well known definitions of the molecular
nanojunction current presented in Refs.\cite{Schreiber06,Fain,Nit7}
\begin{eqnarray}
I=e\frac{d}{dt}\sum_{k\in L}\langle\hat{c}_{k}^{+}\hat{c}_{k}\rangle
=\frac{ie}{\hbar}\sum_{m=1,2}\sum_{k\in L}\langle V_{mk}^{(MK)}\hat{c}_{m}^{+}%
\hat{c}_{k}-h.c.\rangle \nonumber\\
=-\frac{2e}{\hbar}\operatorname{Im}\sum_{m=1,2}%
\sum_{k\in L}V_{mk}^{(MK)}\langle\hat{c}_{m}^{+}\hat{c}_{k}\rangle.\nonumber\\%
\label{eq:I_1}%
\end{eqnarray}
Taking into account that $\langle\hat{c}_{m}^{+}\hat{c}_{k}\rangle=\langle\hat
{c}_{k}^{+}\hat{c}_{m}\rangle^{\ast}$, substituting Eq. (\ref{eq:c^+_kc_m_1})
into Eq. (\ref{eq:I_1}), and using Eqs. (\ref{eq:Term_current}) and
(\ref{eq:GammaMKm}), we have%
\begin{align}
I=-\frac{2e}{\hbar}\operatorname{Im}\sum_{m=1,2}\sum_{k\in L}%
V_{mk}^{(MK)}\langle\hat{c}_{k}^{+}\hat{c}_{m}\rangle^{\ast}\nonumber\\
=e\sum_{m=1,2}\frac{2\pi}{\hbar}\sum_{k\in L}|V_{mk}^{(MK)}|^{2}[n_{m}
-f_{L}(\varepsilon_{k})]\delta(\varepsilon_{k}-\varepsilon_{m})\nonumber\\
=e\sum_{m=1,2}[n_{m}-f_{L}(\varepsilon_{m})]\Gamma_{ML,m}.
\label{eq:I_weak_field}%
\end{align}

\subsection{Calculation of energy transfer}
We now compute the equations of motion for the relaxation
induced by the molecule-metal lead couplings, $\hat{V}_{M}$ and $\hat{V}_{N}$.
We assume that the relaxation processes due to $\hat{V}_{M}$
and $\hat{V}_{N}$ are not independent and also do not depend on the
external magnetic field. Employing a Markovian approximation for the relaxation
induced by the molecule-metal leads coupling, we derive a closed set of
equations for the expectation values of binary function of the electron number
operator in the molecule: $\hat{n}_{m}=\hat{c}_{m}^{+}\hat{c}_{m}$ or $n_{m}\equiv\langle\hat{n}_{m}\rangle$,
where $n_{m}$ is the population of electrons in molecular state $m$. The polarization operator in the molecule is $\hat{b}_{M}=\hat{c}_{m}^{+}\hat{c}_{m^{\prime}}$, $m \neq m^{\prime}$  and the polarization of the molecule is $p_{M}=\langle \hat{b}_{M}\rangle$
and $p_{M}^{+}=\langle \hat{b}_{M}^{+}\rangle$. The molecular excitation population
is $\hat{b}_{k,k^{\prime}}=\hat{c}_{k'}^{+}\hat{c}_{k}$.
From now on, in our derivation we will consider only the HOMO ($|1\rangle$) and LUMO ($|2\rangle$)
among OMO's ($|m\rangle$) and UMO's ($|m^{\prime}\rangle$), respectively, for simplicity, while we may
introduce $|1\rangle=\sum_{m\in M}\rho_{m}|m\rangle$, and $|2\rangle=\sum_{m^{\prime}\in M}\rho_{m^{\prime}}|m^{\prime}\rangle$, where
$\rho_{m}$$(\rho_{m^{\prime}})$ is the density of the electron level $m$$(m')$ of the molecule\cite{Pisani}.
The derivation could be more general so that all
OMOs and UMOs might be considered. In the  case of only HOMO and LUMO, $\hat{c}_{m}$($\hat{c}_{m}^{+}$) and $\hat{c}_{m^{\prime}}$(
 $\hat{c}_{m^{\prime}}^{+}$) will be simply replaced respectively by
$\hat{c}_{1}(\hat{c}_{1}^{+})$ and $\hat{c}_{2}(\hat{c}_{2}^{+})$ variables of the annihilation (creation)
operators for electrons in molecular states $|1\rangle$ and $|2\rangle$ \cite{Fain,Nit05,Nitzan06JCP}.
In the next formulas we shall use the notations made in Ref.\cite{Fain}.
The equations of motion in Eq.(\ref{eq:cm}) and Eq.(\ref{eq:cpm}) include couplings of additional
correlations of the second order $\langle \hat{b}_{mk}\rangle$ due to the electron-transfer interaction $\hat{V}_{M}$ and
higher-order correlations $\hat{b}_{M}\hat{b}_{k,k^{\prime}}$ due to the energy transfer $\hat{V}_{N}$.
Introducing the following notation: $\hat{b}_{M}^{+}=\hat{c}_{2}^{+}\hat{c}_{1}$,
$\text{ }\hat{b}_{M}=\hat{c}_{1}^{+}\hat{c}_{2}$,$\text{ }\hat{b}_{kk^{\prime}}^{+}
=\hat{b}_{k^{\prime}k}=\hat{c}_{k}^{+}\hat{c}_{k^{\prime}}$,
$\text{ }\hat{b}_{kk^{\prime}}=\hat{b}_{k^{\prime}k}^{+}=\hat{c}_{k^{\prime}}^{+}\hat{c}
_{k}$, and taking into account
$\langle \hat{b}_{kk^{\prime}}^{+}\rangle=\langle \hat{b}_{kk^{\prime}}\rangle=f_{k}(\varepsilon_{k})\delta_{kk'}$, we can write:
$\hat{c}_{1}^{+}\hat{c}_{k^{\prime}}^{+}\hat{c}_{k}\hat{c}_{2}=\hat{c}_{1}%
^{+}\hat{c}_{2}\hat{c}_{k^{\prime}}^{+}\hat{c}_{k}=\hat{b}_{M}\hat{b}_{k^{\prime}k}^{+}$.
Then, $V_{N}$ (Eq.(\ref{eq:V_N})) is represented as
\begin{equation}
\hat{V}_{N}=\sum_{K=L,R}\sum_{k^{\prime\prime}\neq k^{\prime\prime\prime}\in
K}(V_{k^{\prime\prime}k^{\prime\prime\prime}}^{(NK)}\hat{b}_{k^{\prime\prime\prime
}k^{\prime\prime}}\hat{b}_{M}^{+}+V_{k^{\prime\prime\prime}k^{\prime\prime}}%
^{(NK)}\hat{b}_{M}\hat{b}_{k^{\prime\prime\prime}k^{\prime\prime}}^{+}).
\end{equation}
From the Heisenberg equation Eq.(\ref{eq:Heis1}), we have
\begin{eqnarray}
\frac{d}{dt}\hat{b}_{M}\hat{b}_{k^{\prime}k}^{+}=\frac{i}{\hbar}[\sum_{m=1,2}%
\varepsilon_{m}\hat{n}_{m}+\sum_{k^{\prime\prime}\in\{L,R\}}\varepsilon%
_{k^{\prime\prime}}\hat{n}_{k^{\prime\prime}},\hat{b}_{M}\hat{b}_{k^{\prime}k}^{+}%
]\nonumber\\%
+\frac{i}{\hbar}[\hat{V}_{N},\hat{b}_{M}\hat{b}_{k^{\prime}k}^{+}].\nonumber\\%
\label{eq:bM}
\end{eqnarray}
The bilinear product of Fermion or pseudo-Fermion operators for $\hat{a}_{i}$ commutes
with the bilinear product of Fermion operators for any function $\hat{f}_{j}$,
 where $i\neq j$. For $i\neq j$, $\lbrack \hat{a}_{i},\hat{f}_{j}]=
\hat{a}_{i}\hat{f}_{j}-\hat{f}_{j}\hat{a}_{i}=0$,
$\lbrack\hat{n}_{m},\hat{b}_{M}\hat{b}_{k^{\prime}k}^{+}]=\hat{n}_{m}\hat{b}_{M}\hat{b}_{k^{\prime}
k}^{+}-\hat{b}_{M}\hat{b}_{k^{\prime}k}^{+}\hat{n}_{m}=[\hat{n}_{m},\hat{b}_{M}]\hat{b}_{k^{\prime}
k}^{+}$, and
$\lbrack\hat{n}_{m},\hat{b}_{M}]=[\hat{c}_{m}^{+}\hat{c}_{m},\hat{c}_{1}^{+}\hat
{c}_{2}]=(-1)^{m-1}\hat{b}_{M}(1-2\hat{n}_{m})$.
For the Poisson brackets for the bilinear products of the Fermion operators, we have
\begin{widetext}
\begin{eqnarray}
\left[\hat{V}_{N},\hat{b}_{M}\hat{b}_{k,k'}^{+}\right]=
-\frac{i}{\hbar}\sum_{K=L,R}\sum_{k\neq k^{\prime}\in K}V_{kk^{\prime}
}^{(NK)}\hat{b}_{k,k^{\prime}}\hat{b}_{k',k}(\hat{b}_{M}
\hat{b}_{M}^{+}-\hat{b}_{M}^{+}\hat{b}_{M})\nonumber\\
=\frac{i}{\hbar}\sum_{K^{\prime}=L,R}\sum_{k^{\prime\prime}\neq
k^{\prime\prime\prime}\in K^{\prime}}V_{k^{\prime}\neq k, k^{\prime\prime}k^{\prime\prime
\prime}}^{(NK)}\hat{b}_{\mathbf{k}^{\prime\prime}}{}_{\mathbf{k}%
^{\prime\prime\prime}}\hat{b}_{k',k}(\sum_{m\neq m'\in M}\hat{n}_{m}-1).
\label{eq:bkk}
\end{eqnarray}
\end{widetext}

\section{Calculation of current}
\label{sec:current}
The simplest approach of transport in a molecular spin nanojunction is to assume that incoming electrons are scattered
both at the noble metal-molecule interfaces and along the molecular chains. Then, the conductance will depend on the net
probability of scattering \cite{Nit10}. Elastic scattering does not forbid electrons to transport through the nanojunction.
The coherent conductance takes place in most molecular chains and nanowires when the electron transport occurs far from a resonance frequency
between the metal Fermi energy and the molecular eigenstates at low temperatures\cite{Nit10}.
Landauer theory assumes that electrons move smoothly from one electrode ($L$) to another ($R$)
only by elastic scattering within nanojunction. In the presence of magnetic impurities or ferromagnetic leads,
electrons would show the spin dependent transport ( ''spin valve behavior'') in the presence of magnetic field. The spin polarized
electron current emission is excited by magnetic field between metal leads.
Thus, we take into account the linear form of new spin electron polarization added in the Landauer formalism.
The total current $I$ is taken by the rate of change of occupation number
operator of electrons in the molecule is described in \cite{Schreiber06,Fain,Nit7}.
In Eq.(\ref{eq:I_1}), the current
$I$ represents the rate of flow of electrons from the left electrode to the molecule.
thus using the previous results for $\langle c_{k}^{+}c_{m}\rangle$ we obtain%
\begin{widetext}
\begin{eqnarray}
\frac{d\hat{c}_{k}^{+}}{dt}\hat{c}_{m}=\frac{i}{\hbar}\varepsilon_{k}%
\hat{c}_{k}^{+}\hat{c}_{m}+\frac{i}{\hbar}\sum_{K=L,R}\sum_{k\in K}%
V_{km}^{(MK)}\hat{c}_{k}^{+}\hat{c}_{m}\nonumber\\
+\frac{i}{\hbar}\sum_{K=L,R}\sum_{k\neq k^{\prime}\in K}\{V_{k^{\prime}%
k}^{(NK)}\hat{c}_{1}^{+}\hat{c}_{k^{\prime}}^{+}\hat{c}_{k}\delta_{2m}\hat
{c}_{m}+V_{kk^{\prime}}^{(NK)}\hat{c}_{k}^{+}\hat{c}_{k^{\prime}}\hat{c}%
_{2}^{+}\delta_{1m}\hat{c}_{m}\},
\label{eq:dcm+cm}
\end{eqnarray}
\end{widetext}
\begin{widetext}
\begin{align}
\hat{c}_{k}^{+}\frac{d\hat{c}_{m}}{dt}  &  =-\frac{i}{\hbar}\varepsilon
_{m}\hat{c}_{k}^{+}\hat{c}_{m}-\frac{i}{\hbar}\sum_{K=L,R}\sum_{k\in K}%
V_{mk}^{(MK)}\hat{c}_{m}^{+}\hat{c}_{k}+\frac{i}{\hbar}\mathbf{H}_{B}
(\mathbf{r},t)\hat{c}_{k}^{+}\hat{c}_{m'\neq m}-\nonumber\\
&  -\frac{i}{\hbar}\sum_{K=L,R}\sum_{k\neq k^{\prime}\in K}\{V_{kk^{\prime}%
}^{(NK)}\delta_{2m}\hat{c}_{m}^{+}\hat{c}_{k}^{+}\hat{c}_{k^{\prime}}\hat
{c}_{1}+V_{k^{\prime}k}^{(NK)}\delta_{1m}\hat{c}_{m}^{+}\hat{c}_{2}\hat
{c}_{k^{\prime}}^{+}\hat{c}_{k}\}
-\frac{i}{\hbar}\sum_{m\neq m^{\prime}\in M}V_{mm^{\prime}}
^{(VK)}(\hat{c}_{k}^{+}\hat{Q}_{\alpha}^{a}\hat{c}_{m})\text{.}
\label{eq:cm+dcm}
\end{align}
\end{widetext}
Summing Eq.(\ref{eq:dcm+cm}) and (\ref{eq:cm+dcm}) and taking the expectation
values, we have:
\begin{widetext}
\begin{eqnarray}
\frac{d}{dt}\langle\hat{c}_{k}^{+}\hat{c}_{m}\rangle=\frac{i}{\hbar
}(\varepsilon_{k}-\varepsilon_{m})\langle\hat{c}_{k}^{+}\hat{c}_{m}
\rangle+\frac{i}{\hbar}\sum_{m^{\prime}\in M}V_{m^{\prime}k}^{(MK)}\langle
\hat{c}_{m^{\prime}}^{+}\hat{c}_{m}\rangle-\frac{i}{\hbar}\sum_{K=L,R}
\sum_{k^{\prime}\in K}V_{mk^{\prime}}^{(MK)}\langle\hat{c}_{k}^{+}\hat
{c}_{k^{\prime}}\rangle+\nonumber\\
+\frac{i}{\hbar}\sum_{k^{\prime}\neq k\in K}V_{k^{\prime}k}^{(NK)}
\langle\hat{c}_{k^{\prime}}^{+}(\hat{c}_{2}^{+}\hat{c}_{1}+\hat{c}_{1}^{+}
\hat{c}_{2})\hat{c}_{m}\rangle+\frac{i}{\hbar}\mathbf{H}_{B}
(\mathbf{r},t)\langle\hat{c}_{k}^{+}\hat{c}_{m}\rangle-\nonumber\\
-\frac{i}{\hbar}\sum_{K=L,R}\sum_{k^{\prime\prime}\neq k^{\prime}\in
K}\langle V_{k^{\prime\prime}k^{\prime}}^{(NK)}\hat{c}_{k}^{+}\hat
{c}_{k^{\prime\prime}}^{+}\hat{c}_{k^{\prime}}\hat{c}_{1}\delta_{2m}
+V_{k^{\prime}k^{\prime\prime}}^{(NK)}\delta_{1m}\hat{c}_{k}^{+}\hat{c}
_{2}\hat{c}_{k^{\prime}}^{+}\hat{c}_{k^{\prime\prime}}\rangle
-\frac{i}{\hbar}\sum_{m\neq m'\in M}\sum_{k\in L}V_{mm'}^{VK}(
\hat{c}_{k}^{+}\hat{Q}_{\alpha}^{a}\hat{c}_{m})\text{.}
\label{eq:cmcurr}
\end{eqnarray}
\end{widetext}
for the current we have from Eq.(\ref{eq:I_1})
and by full analogy with Eqs. (\ref{eq:cm+dcm}) and (\ref{eq:cmcurr}),
we obtain a general formula for the total current given by the
rate of change of the occupation number operator of electrons in
the molecule.
\begin{eqnarray}
I=\frac{ie}{\hbar}\sum_{m^{\prime}\in M}\sum_{k\in L}\langle
V_{m^{\prime}k}^{(MK)}\hat{c}_{m^{\prime}}^{+}\hat{c}_{k}-h.c.\rangle+\nonumber\\
+\frac{ie}{\hbar}\{\sum_{k\in L}\sum_{k^{\prime}\neq k\in L}\langle
V_{k^{\prime}k}^{(NK)}\hat{c}_{k^{\prime}}^{+}(\hat{c}_{2}^{+}\hat{c}_{1}
+\hat{c}_{1}^{+}\hat{c}_{2})\hat{c}_{k}\rangle\nonumber\\
-\sum_{k\in L}\sum_{k^{\prime
}\neq k\in L}\langle V_{kk^{\prime}}^{(NK)}\hat{c}_{k}^{+}(\hat{c}_{2}^{+}
\hat{c}_{1}+\hat{c}_{1}^{+}\hat{c}_{2})\hat{c}_{k^{\prime}}\rangle\}\nonumber\\
+\frac{i}{\hbar}\sum_{k\in L}\mathbf{H}_{B}
(\mathbf{r},t)\langle\hat{c}_{k}^{+}\hat{c}_{m}\rangle\nonumber\\
-\frac{ie}{\hbar}\sum_{m\neq m'\in M}\sum_{k\in L}V_{mm'}^{VK}(\hat{c}_{k}\hat{Q}_{\alpha}^{a}\hat{c}_{m}^{+}+
\hat{c}_{k}^{+}\hat{Q}_{\alpha}^{a}\hat{c}_{m}).\nonumber\\
\label{eq:curr}
\end{eqnarray}
Here, the energy transfer, the electron-phonon interaction
as well as the external magnetic field effect is included.
The term in the braces on the right-hand side of the last equation is equal to
zero. This can be seen if we exchange $k\leftrightarrows k^{\prime}$ in the
second term of the braces.
From Eq.(\ref{eq:curr}) one would select the part of current excited by magnetic field by using
Eqs. (\ref{eq:c^+_kc_m_1}-\ref{eq:I_weak_field}). We have a simplified form for the current:
\begin{equation}
I_{H}=\frac{ie}{\hbar}H_{B}(r,t)\sum_{m\in M}[n_{m}-f_{K}(\varepsilon_{k})]\Gamma_{ML,m}.
\end{equation}
For the Hamiltonian of magnetic field and spin-spin interaction $\hat{H}_{S}$ (Eq.(\ref{eq:Hs})), we have
\begin{equation}
I_{S}=-\frac{2ie}{\hbar}\sum_{m>m',\sigma\sigma'}J_{m,m'}S_{m}S_{m'}
[n_{m}-f_{K}(\varepsilon_{k})]\Gamma_{ML,m}.
\end{equation}
It means that there is a spontaneous current due to the transient process of spin-spin interactions with
the fast damping defined by the coefficient $\Gamma_{ML,m}$.
The transient process of spontaneous spin-spin interaction current can occur in the molecular structure both in the presence and absence of any external field. The charge transferred during the time of
an electromagnetic pulse with the  finite duration is given by
$Q=\int_{-\infty}^{\infty}I(t)dt$. In the next section, we shall provide the computation of the electric current
with $\langle\hat{c}_{k}\hat{Q}_{\alpha}^{a}\hat{c}_{m}^{+}\rangle$ and
$\langle\hat{c}_{k}^{+}\hat{Q}_{\alpha}^{a}\hat{c}_{m}\rangle$.

\section{Differential equations for nonequilibrium charge transfer}
\label{sec:diffeq}
In the form of Markovian approximation for the relaxation
induced by the molecule-metal leads coupling for the dynamics of the electron system of the
molecular junction, we have derived a closed set of
equations for the expectation values of binary operator of $\langle\hat{n}_{m}%
\rangle=n_{m}$ and $\langle\hat{b}_{M}\rangle=p_{M}$, which are variables of the annihilation and creation
operators for electrons in molecular states $|1\rangle$ and $|2\rangle$.
Straightforward operator algebra manipulations yield $n_{m}$ and $p_{M}$
in the rotating wave approximations. The expression for nanojunction in the Hartree-Fock approximation
$n_{1}n_{2}+\langle{p}_{M}\rangle\langle
{p}_{M}^{+}\rangle=\frac{1}{2}(n_{1}+n_{2})$, is correct
only when the length of the Bloch vector is conserved. However, due to the
charge transfer between the molecular orbitals and metals, this value is not
conserved. Thus, the so-called relevant density matrix of molecule $\rho_{M}$ is used as a
total density matrix $\rho$, which contains the information of the expectation
values of operators $\hat{n}_{m}$ and $\hat{p}_{M}$. If chosen appropriately, the
relevant density matrix contains the essential part of the molecular dynamics, but we employ the following properties
\cite{Koch,Grab}:
$Tr(\hat{n}_{1}\hat{n}_{2}\rho_{M})=|p_{M}|^{2}+n_{1}n_{2}\label{eq:n_eMxn_hM}$.
Due to the assumption mentioned above, the scattering rate of populations depends on
the polarization $\hat{p}_{M}$ which is similar to the case of semiconductor Bloch equations \cite{Koch}.
It is assumed that the polarization is small, so that in the final scattering terms
they keep only the terms linear to the polarization \cite{Koch}.
The advantage of our approach to the molecular junction is that it could
also be used when the dissipative system (metals) is not in equilibrium and the
many-body effects are significant. In this case, the total density matrix is $\rho=\rho_{M}%
{\displaystyle\prod\limits_{k\in L,R}}
\rho_{k}$.
By taking into account the formal mathematical methods, the derivation of the differential equations for the
expectation values from the Heisenberg equations are reported\cite{Koch,Fain}.
Employing Eqs.(\ref{eq:bM}), (\ref{eq:bkk}) and (\ref{eq:cmcurr}) as well as (\ref{eq:dc^+_kc_1}),
(\ref{eq:dc^+_kc_2}) and (\ref{eq:dc^+_kc_1int}) in Appendix \ref{sec:AppD}
for the polarization, we finally obtain
\begin{eqnarray}
\frac{dp_{M}}{dt}=-i[\omega_{0}+\sum_{K=L,R}\Delta_{MK}-\omega
(t)]p_{M}\nonumber\\
-\frac{i}{2}\Omega_{R}(\sum_{m\in M}n_{m}-1)-p_{M}\Gamma_{PM}
\label{eq:dpm}
\end{eqnarray}
where $\omega(t)=\omega_{0}+\mu (t-t_{0})$, $\mu$ is a chirp rate,
\begin{equation}
\Gamma_{PM}=\sum_{K=L,R}\{\frac{1}{2}(\Gamma%
_{MK,1}+\Gamma_{MK,2})+\Gamma_{NK}[\omega(t)]\}
\end{equation}
and
\begin{equation}
\Delta_{MK}=\frac{1}{\hbar}P\sum_{k\in K}[\frac{|V_{k1}^{(MK)}|^{2}%
}{\varepsilon_{k}-\varepsilon_{1}}-\frac{|V_{2k}^{(MK)}|^{2}}{\varepsilon
_{k}-\varepsilon_{2}}] \label{eq:Im(MKterm_for_pM)}.%
\end{equation}
The $\Omega_{R}$ is the generalized Rabi frequency\cite{Rabi1} consisting of two parts.
\begin{eqnarray}
\Omega_{R}=g\mu_{B}\hbar^{-1}\mathcal{B}(t)\sum_{m,m'\in M}S_{m}\nonumber\\
-2\hbar^{-1}\sum_{m>m',\sigma\sigma'}J_{m,m'}S_{m}S_{m'}.
\label{eq:omr}
\end{eqnarray}
The first part is the Larmor frequency
due to nonstationary magnetic fields and the second part is the
spin-spin interaction. The equation for the electron number has the form:
\begin{widetext}
\begin{eqnarray}
\frac{dn_{m}}{dt}=\Omega_{R}\operatorname{Im}p_{M}+\sum
_{K=L,R}[f_{mK}(\hbar\omega_{0}/2)-n_{m}]\Gamma_{MK,m}\nonumber\\
+2\sum_{m\neq m'\in M}(n_{m}-f_{K}(\varepsilon_{k}))(\overline{n}_{Qm}+n_{Qm})\Gamma_{VK,m}\nonumber\\
-\sum_{K=L,R}\{\Gamma_{NK}(\omega_{0})[|p_{M}
|^{2}+\prod_{m\in M}n_{m}]-\Gamma_{NK}(\omega_{0})[1-\sum_{m\in M}n_{m}
]\}.\label{eq:n_eM2}
\end{eqnarray}
\end{widetext}
Here we introduce the next basic dynamical variables for the phonon number which is
$n_{ph}\equiv\langle\hat{a}_{\alpha}^{+}\hat{a}_{\alpha}\rangle$, and the phonon-assisted density matrices
of electron-phonon interaction are $n_{Qm}\equiv\langle\hat{c}_{k}\hat{Q}_{\alpha}^{a}\hat{c}_{m}^{+}\rangle$,
$\overline{n}_{Qm}\equiv\langle\hat{c}_{k}^{+}\hat{Q}_{\alpha}^{a}\hat{c}_{m}\rangle$,
$n_{Pm}\equiv\langle\hat{c}_{k}\hat{P}_{\alpha}^{a}\hat{c}_{m}^{+}\rangle$,
$\overline{n}_{Pm}\equiv\langle\hat{c}_{k}^{+}\hat{P}_{\alpha}^{a}\hat{c}_{m}\rangle$,
$Q_{\alpha}\equiv\langle \hat{Q}_{\alpha}^{a}\rangle$,
$P_{\alpha}\equiv\langle \hat{P}_{\alpha}^{a}\rangle$ where
$\hat{P}_{\alpha}^{a}=i(\hat{a}_{\alpha}^{+}-\hat{a}_{\alpha})$.
Within the Heisenberg picture, we have a set of differential equations
for the phonon occupation number $n_{ph}$ and the expectation values of $n_{Qm}$ and $\overline{n}_{Qm}$ derived in refs.\cite{Haug2,Shilp1,Shilp2,buts}.
Substituting Eqs.(\ref{eq:cQc+}, \ref{eq:c+Qc} and
\ref{eq:cQc+M}, \ref{eq:cQc+E}) into Heisenberg equations Eqs.(\ref{eq:Heis1},\ref{eq:Heis2}) and
using Pauli commutators and anticommutators  Eqs. (\ref{eq:anticommutation1}, \ref{eq:anticommutation2})
for the phonon-electron interaction terms, we finally obtain
\begin{eqnarray}
 \frac{dn_{Qm}}{dt}=\omega_{\alpha}n_{Pm}\nonumber\\
+2i(1-n_{m})Q_{\alpha}\Gamma_{QM,m}
-2i n_{m}Q_{\alpha}\overline{\Gamma}_{QM,m}\nonumber\\
+(2n_{ph}+1)(1-2n_{m})(n_{m}-f_{K}(\varepsilon_{k}))\Gamma_{VK,m}
\label{eq:n_se}
\end{eqnarray}
\begin{eqnarray}
\frac{d\overline{n}_{Qm}}{dt}=\omega_{\alpha}\overline{n}_{Pm}+
i(\omega_{k}-\omega_{m})\overline{n}_{Qm}\nonumber\\
+2i(\omega_{m}n_{m}-\omega_{k}f_{K}(\varepsilon_{k}))
\overline{n}_{Qm}\nonumber\\
+(2n_{ph}+1)(2n_{m}+1)(n_{m}-f_{K}(\varepsilon_{k}))\Gamma_{VK,m}
\label{eq:n_sh}
\end{eqnarray}
\begin{eqnarray}
\frac{dn_{Pm}}{dt}=-\omega_{\alpha}n_{Qm}
+2i(1-n_{m})P_{\alpha}\Gamma_{QM,m}\nonumber\\
-2i n_{m}P_{\alpha}\overline{\Gamma}_{QM,m}
-2(n_{m}-f_{K}(\varepsilon_{k}))\Gamma_{VK,m}
\label{eq:p_se}
\end{eqnarray}
\begin{eqnarray}
\frac{d\overline{n}_{Pm}}{dt}=-\omega_{\alpha}\overline{n}_{Qm}
+i(\omega_{k}-\omega_{m})\overline{n}_{Pm}\nonumber\\
+2i(\omega_{m}n_{m}-\omega_{k}f_{K}(\varepsilon_{k}))
\overline{n}_{Pm}\nonumber\\
-2(n_{m}-f_{K}(\varepsilon_{k}))\Gamma_{VK,m}
\label{eq:p_sh}
\end{eqnarray}
where $\omega_{k}=\varepsilon_{k}/\hbar$, $\omega_{m}=\varepsilon_{m}/\hbar$, and
\begin{equation}
\frac{dQ_{\alpha}}{dt}=\omega_{\alpha}P_{\alpha}
\label{eq:Q_sh}
\end{equation}
\begin{equation}
\frac{dP_{\alpha}}{dt}=-\omega_{\alpha}Q_{\alpha}+\sum_{K=L,R}(n_{m}-f_{K}(\varepsilon_{k}))\Gamma_{MK,m}.
\label{eq:P_sh}
\end{equation}
The equation for $n_{ph}$ is
\begin{eqnarray}
\frac{dn_{ph}}{dt}=\sum_{m\neq m'\in M}(n_{m}-f_{K}(\varepsilon_{k}))\overline{n}_{Pm}\Gamma_{VK,m}\nonumber\\
+\sum_{m\neq m'\in M}n_{Pm}\Gamma_{QM,m}.
\label{eq:n_ph}
\end{eqnarray}
Here
\begin{equation}
\Gamma_{QK,m}=\frac{1}{\hbar}\sum_{k\in K}V_{k,m}^{(VK)}f_{K}(\varepsilon_{k}),
\end{equation}
\begin{equation}
\overline{\Gamma}_{QK,m}=\frac{1}{\hbar}\sum_{k\in K}V_{k,m}^{(VK)}(1-f_{K}(\varepsilon_{k})),
\end{equation}
\begin{equation}
\Gamma_{VK,m}=
\frac{\pi}{\hbar}\sum_{m\neq m'\in M}\sum_{k\in K}V_{m,k}^{(MK)}V_{m,m'}^{(VK)}f_{K}(\varepsilon_{k})\delta(\varepsilon_{k}-\varepsilon_{m}),
\end{equation}
\begin{widetext}
\begin{equation}
\Gamma_{NK}(\omega(t))=\frac{2\pi}{\hbar}\sum_{k,k'\in K}
|V_{k,k'}^{(NK)}|^{2}f_{K}(\varepsilon_{k})f_{K}(\hbar\omega-\varepsilon_{k})\delta(\hbar\omega-\varepsilon_{k}).
\end{equation}
\end{widetext}
From Eq.(\ref{eq:curr}) we have a part of the full formula for the current taking into account the electron-phonon interaction:
\begin{equation}
I_{ph}=-e\sum_{m\neq m'\in M}(n_{Qm}\overline{\Gamma}_{QK,m}+\overline{n}_{QK,m}\Gamma_{QK,m})
\end{equation}

\section{magnetic control of current and transferred charge with chirped
pulses}

\label{sec:Magnetcont}

Now we have generalized theoretical results for the charge transfer
at the quasi-stationary strong magnetic field limit, as discussed  in the previous section.
The well-known procedures are based on the coherent excitation which
produces the complete population inversion in an ensemble of degenerate two-level molecules
with the Rabi population oscillations by the Gaussian pulse excitation \cite{All75}.
It has been demonstrated that a molecule or an atom excited by the Gaussian pulses behaves as a semiconductor quantum
dot\cite{Stievater01,Kamada_Takagahara01,Htoon_Takagahara02,Zrenner02Nature}.
To solve the main problem of the Gaussian pulse excitation of molecular levels, it requires the information of the
resonant magnetic source, the precise control
of the pulse area, and the chirp rate $\mu$ \cite{Sho92}.
In order to provide the complete population inversion procedure, known as adiabatic rapid passage (ARP)
\cite{Mel94,Sho92,All75,Tre68,Vit01,Fai04JCP,Fai05JOSAB},
the entire population needs to transform from ground $|1>$ to the excited $|2>$
electronic state. Thus, it is necessary to sweep the pulse frequency through a
resonance. The mechanism of ARP can be explained by avoided crossing of
dressed (adiabatic) states. In particular, starting from state $|1>$, the system follows the adiabatic state
and eventually ends up in state $|2>$ \cite{Vit01}. The scheme
based on ARP is robust since it is insensitive to the pulse area and the precise
location of the resonance. Therefore, we shall focus on the following ARP procedures
as a way to control the magnetic field induced charge transfer in molecular spin nanojunctions.
Our formalism presented for the coherent magneto-spin properties of
nanojunctions of molecular and quantum dots by the Gaussian pulses excitation
\cite{Zrenner02Nature}, is analyzed here.

\begin{figure}
[ptb]
\begin{center}
\includegraphics[width=3.4in]%
{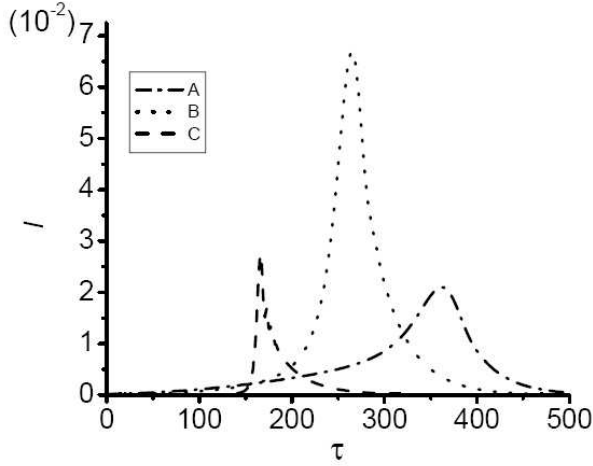}
\caption{Current $I$ (in dimensionless unit) as a
function of time $\tau=\omega_{0}t$ for the linear chirp rate ${\mu
}/\omega_{0}^{2}=5\cdot10^{-3}$ (A)$,$ $10^{-2}$ (B), and $10^{-1}$ (C).
The parameters of the calculation are as follows
$\delta=5\cdot10^{-3}$, $\hbar\omega_{0}=3$ eV,
$\Gamma_{MR,1}/\hbar\omega_{0}=0.04$,
$\Gamma_{MR,2}/\hbar\omega_{0}=0.03$,
$\Gamma_{NK}/\hbar\omega_{0}=0.01$, $\Gamma_{QM,m}=0.01$, $\overline{\Gamma}_{QM,m}=0.99$.
$d_{\mu}\mathcal{B}_{0}/\hbar\omega_{0}=0.2$, and $\delta/\omega_{0}=0.1$,
The picture illustrates how
the current depends on the linear chirp rate.}
\label{fig:chirp_mu_a}
\end{center}
\end{figure}

\begin{figure}
[ptb]
\begin{center}
\includegraphics[width=3.4in]
{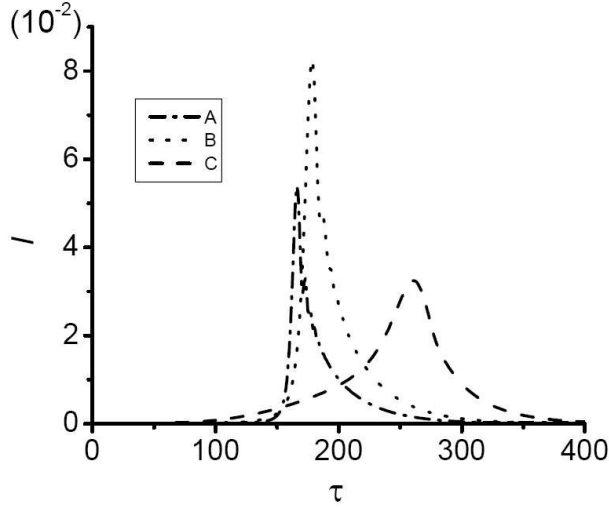}
\caption{Current $I$ (in dimensionless unit) as a function of time $\tau=\omega_{0}t$
for the linear chirp rate $\mu/\omega_{0}^{2}=10^{-1}$ (A)$,$ $5\cdot10^{-2}$
(B) and $10^{-2}$ (C), where $\delta=10^{-2}$. Other parameters are identical to those of
Fig.\ref{fig:chirp_mu_a}. The picture illustrates how signatures of
ARP increase the amplitude of the induced current.}
\label{fig:chirp_curr_b}
\end{center}
\end{figure}

\begin{center}
\begin{figure}
\includegraphics[width=3.4in]{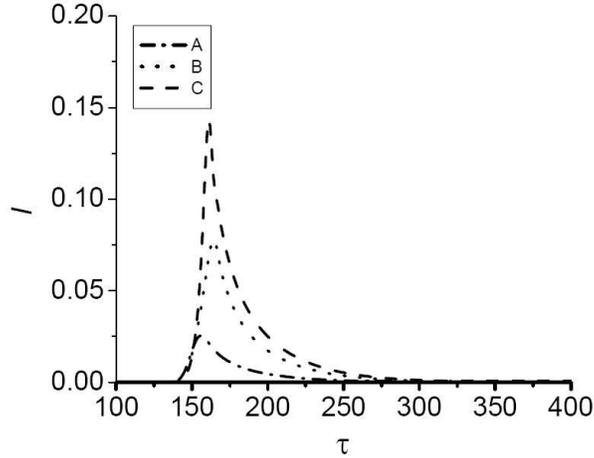}
\caption{ The current $I$ (in dimensionless unit) transferred after the completion of the pulse
action as a function of the chirp rate $\mu$ in the frequency domain,
 $\tau=\omega_{0}t$. Here $d_{\mu}\mathcal{B}_{0}/\hbar\omega_{0}=0.2$, $\delta=10^{-1}$,
$\mu/\omega_{0}^{2}=10^{-2}$ (A)$,$ $5\cdot10^{-2}$
(B) and $1.5\cdot10^{-1}$ (C) for the transform-limited pulse. In the course of chirping,
the pulse energy is conserved so that $\int_{-\infty}^{\infty
}\mathcal{B}^{2}(t)dt=\mathcal{B}_{0}^{2}$ is constant. Other
parameters are identical to those of Fig.\ref{fig:chirp_mu_a}.
The picture illustrates the manifestation of ARP, i.e. how the amplitude of the current
increases with increasing chirp rate $\mu$.}
\label{fig:chirp_Fi}
\end{figure}
\end{center}

As a particular example, we shall consider a magnetic-induced charge transfer in
molecular nanojunctions. The instantaneous
magnetic pulse frequency $\omega(t)$ is given as the linear chirped pulses $\omega(t)=\omega_{0}
-\mu(t-t_{0})$ (where chirped rate $\mu$ is constant) during the pulse excitation\cite{Sho92}, and
the Gaussian pulse of the magnetic field is used as\cite{All75,Sho92,Fai98}
\begin{equation}
B(t)\equiv \mathcal{B}_{0}\exp[-\frac{1}{2}(\delta^{2}-i\mu)(t-t_{0}%
)^{2}], \label{eq:gausspulse}
\end{equation}
where $\delta$ is the inverse duration of the pulse and $\mathcal{B}_{0}$ is the amplitude of pulse.
$\delta$ and $\mu$ are used in dimensionless
units below. Particularly interesting in this respect are molecules
characterized by strong charge-transfer transitions that are
reflected in the formation of an excited molecular state with a
magnetic dipole. When the magnetic fields operate as a molecular dipole connecting
two metal leads along the direction of the charge transfer
(approximately perpendicular to the current flow axis), the magnetic pumping
into the charge-transfer state creates an internal driving force
for charge flow between the two leads. Here we will make a reasonable assumption that a
charge-transfer transition within the molecule is expressed in
terms of the change in relative coupling strengths of the molecular
HOMO and LUMO to their metallic contacts.
We thus investigate
models in which $\Gamma_{MK,1}<\Gamma_{MK,2}$, $K\in L,R$. This
inequality reflects the fact that the excited molecular state is
dominated by atomic orbitals of larger amplitude on one side
of the molecule than on the other side, resulting in greater
overlap with metal orbitals on that side. In the calculation we used the next constants\cite{Fain,Nitzan06JCP}:
$\Gamma_{ML,1}=0.01$ eV, $\Gamma_{ML,2}=0.02$ eV and $\Gamma_{VK,m}=0.02$ eV.
The numbers taken above for the $\Gamma_{MK,m}$ parameters where
$K=\{L,R\}$ and $m=1,2$ are reasonable, and in any case
we find that similar results are obtained when they are
changed within a reasonable range. Also, the choice $0.01<\Gamma_{VK,m}<0.1$
$eV$ reflects an assumed lifetime of $\approx10$ fs for an
excited molecule at the metal surface to relax via the
electron-phonon mechanism, which is also a reasonable number.
The energy of the pulse can be evaluated
as 1-10 eV \cite{Schreiber06}.
This number is the order of
magnitude of normal magnetic field intensities used in spectroscopy,
and it should be kept in mind that it could result from
weaker incident fields due to local field enhancement. Figs. \ref{fig:chirp_mu_a}-\ref{fig:chirp_Fi} show the influence of
$\mu$ or the chirp rate in the time domain on the induced current calculated by Eq.(\ref{eq:curr}) in dimensionless form:
$I=I(\tau)/e\Gamma_{ML,1}$ during one magnetic pulse action.
These results are displayed as a function of dimensionless time $\tau=\omega_{0}t$
with the $\mu$ dependence from the numerical solution of Eqs. (\ref{eq:dpm}-\ref{eq:n_ph})
 for a Gaussian pulse. We see that the pulse chirping can
increase the amplitudes of the induced
current, which can be explained by signatures of
ARP (Fig.\ref{fig:chirp_mu_a}),(Fig.\ref{fig:chirp_curr_b}).

The pulses of the current obtained by changing the separation of pulse
compression has to be Lorentzian. The parameter $\delta$ is the inverse pulse
duration of the corresponding transform-limited pulse. The chirped frequency  $\omega(t)$
changes to the resonance condition during the time of the pulse.
Note that the local field of Eqs. (\ref{eq:hb1})
and (\ref{eq:hb2}) in the nanojunction also reflects mangnon
excitation in the leads. The incident pulse shape
affected only by the compression uses the possible contribution of the near-field
response to both plasmonic and magnon excitations in the leads and to excitons in the molecule.
The excitation of electric current by the magnetic fields in the
spin nanojunction also reflects phonon-electron interaction in the molecule and the leads.
The incident pulse shape affected only by the pulse compression excites
the phonon-electron interaction toward the broadening of the peak of the current
response \cite{Wang_Shen06,Brixner06}. Such effects are presented in Figs. \ref{fig:chirp_mu_a}-\ref{fig:chirp_Fi}.
The observable of interest is the magnetic field induced electronic current.
Changing bias under magnetic fields with a fixed frequency
can make the polar molecule into and out of the resonance by excitation,
leading to highly nonlinear current voltage dependence
including the possibility for negative differential
resistance.
\begin{figure}
[ptb]
\begin{center}
\includegraphics[width=3.4in]{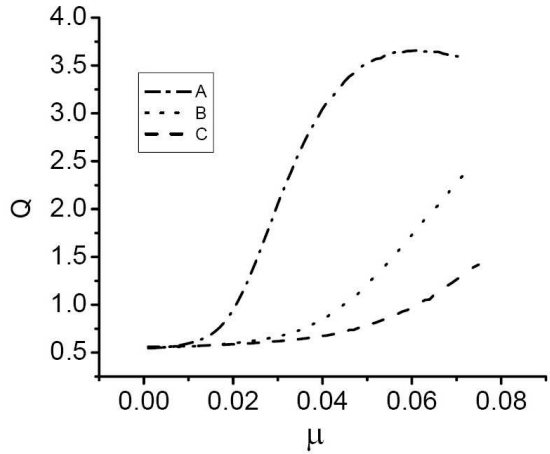}
\caption{Charge $Q$ (in dimensionless unit) transferred after the completion of the pulse
action as a function of the chirp rate $\mu$ for the transform-limited pulse. Here,
$d_{\mu}\mathcal{B}_{0}/\hbar\omega_{0}=0.2$, $\delta=10^{-3}$ (A), $\delta=5\cdot10^{-3}$ (B), and
$\delta=10^{-2}$ (C). Other parameters are identical to those of Fig.\ref{fig:chirp_mu_a}.
The picture illustrates the manifestation of ARP, i.e., how the quantity of the charge
increases with increasing chirp rate $\mu$.}
\label{fig:chirp_d}
\end{center}
\end{figure}
\begin{figure}
[ptb]
\begin{center}
\includegraphics[width=3.4in]{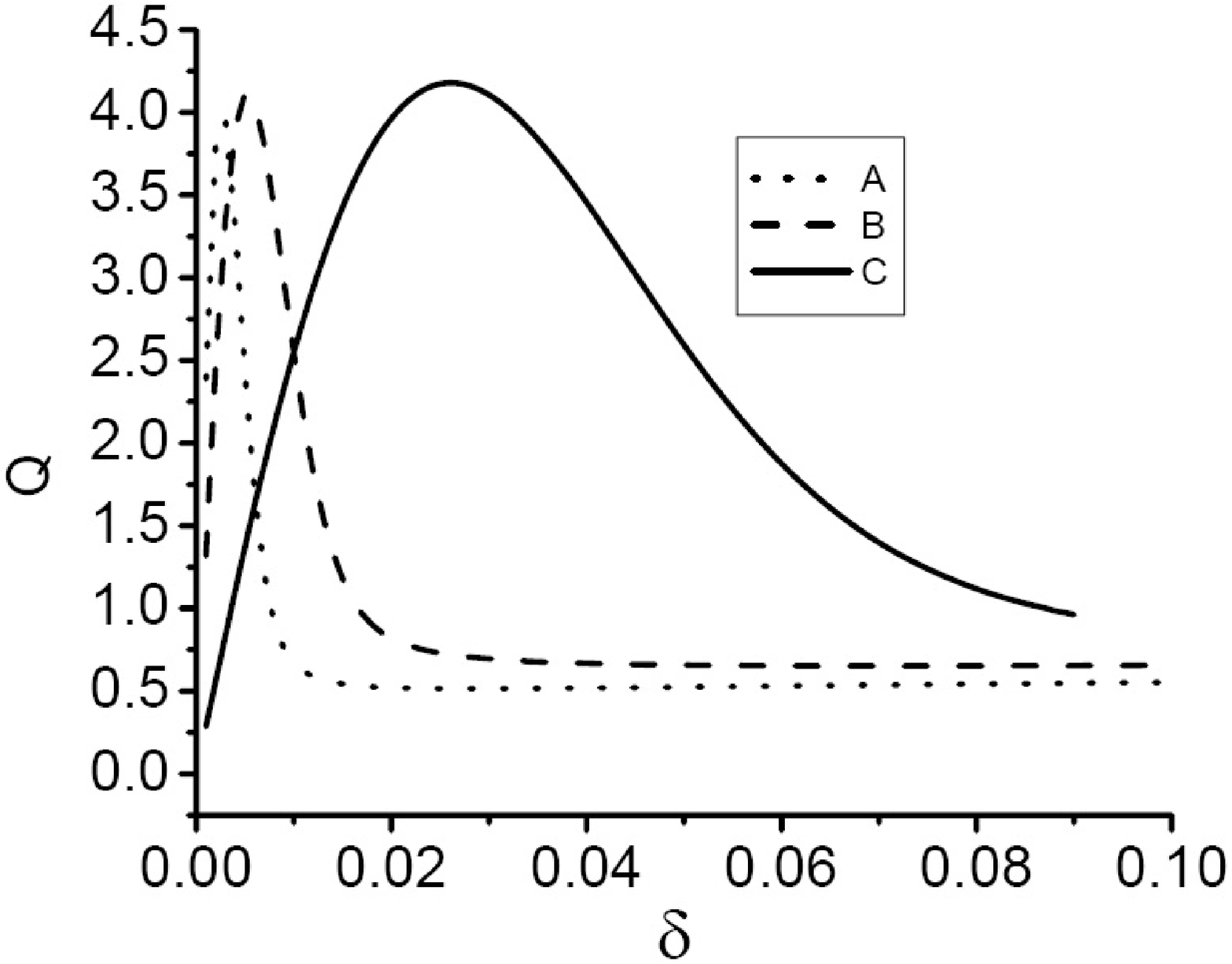}
\caption{Charge $Q$ (in dimensionless unit) transferred after the completion of the pulse
action as a function of $\delta$ in the presence of energy
transfer. Chirp rate is $\mu=5\cdot10^{-3}$ (A), $\mu=10^{-2}$ (B), and
 $\mu=5\cdot10^{-2}$ (A). Other parameters are identical to
those of Fig.\ref{fig:chirp_Fi}. }%
\label{fig:chirp_Fi_4corr}%
\end{center}
\end{figure}
Figs. \ref{fig:chirp_d} and \ref{fig:chirp_Fi_4corr} show the calculation
results of the transferred charge $Q$:
\begin{equation}
Q=\int_{-\infty}^{\infty}I(\tau)d\tau
\end{equation}
as a function of the chirp rate $\mu$ in the
frequency domain and $\delta$. The calculated dependencies
$Q(\mu)$ of Fig.\ref{fig:chirp_d} and $Q(\delta)$ of Fig.\ref{fig:chirp_Fi_4corr} are
confined to the values of arguments $\mu$ and $\delta$ corresponding to $d_{\mu}\mathcal{B}_{0}/\hbar\omega
_{0}\leq0.3$ ($d_{\mu}=g\mu_{B}/2$ is the molecular magnetic dipole moment, Eq.(\ref{eq:hb2})).
Since the theory uses the rotating wave approximation, the amplitude of magnetic field has to be limited.
One can see that $Q$ grows rapidly for small
$\mu$ and $\delta$. Fig.\ref{fig:chirp_d} shows that the growth
of $Q$ is slow for moderate $\delta=10^{-2}$ and then $Q$ tends to become a constant value for large
$\delta=10^{-3}$. The large pulse energy is larger than the value of $Q$ at which the growth of $\delta$ slows down as presented in
Figs. \ref{fig:chirp_d} and \ref{fig:chirp_Fi_4corr}, which illustrate the influence of the energy transfer that diminishes the
corresponding values of $Q$ (see also Figs. \ref{fig:chirp_mu_a} and \ref{fig:chirp_curr_b}).
Figs. \ref{fig:chirp_d} and \ref{fig:chirp_Fi_4corr} show the resulting behavior of the charge current induced
by magnetic fields from Eq.(\ref{eq:gausspulse}), based on the full self-consistent calculation described in Sec. II. The
parameters used in this calculation are $T=300$ $K$.
As expected, a steady state current flows through the
spin nanojunction in the presence of magnetic pumping.
A peak of the current occurs at the frequency of the charge-transfer transition,
i.e., the HOMO-LUMO energy gap in our model.
The fact that magnetic current can occur in a molecular spin nanojunction
with the postulated characteristics is a direct consequence
of the fact that the charge-transfer properties of the
molecules lead to an internal driving force that would result in
magnetic voltage in the corresponding open circuit.
Another point of concern is the thermal stability of spin nanojunction under the proposed
thermal heating. On the other hand, the current calculated with
these parameters (Figs.\ref{fig:chirp_mu_a}-\ref{fig:chirp_Fi}) is of order $1$ nA, implying
that the magnetic intensity which is lower by an order of magnitude
can still lead to observable currents. We conclude that the magnetic current
in a molecular spin nanojunction is a realistic possibility.
The values of $Q$ shown in Fig. \ref{fig:chirp_Fi_4corr}
can be rationalized by the theoretical consideration below. It illustrates the influence of $\delta$ of magnetic pulse,
the carrier pulse frequency $\omega$, and the corrected frequency of the
molecular transition $\omega_{0}$ on the transferred charge $Q$. As shown in Fig.\ref{fig:chirp_Fi_4corr},
there is an optimal parameter for magnetic pulses to provide the maximum of $Q$. This picture shows us the
solution to the optimal control of the parameters of the pulse in
order to obtain the maximum charge transfer.
To end this section we note that the current corresponding to the
expectation value of $Q$ is $0.7\cdot10^{-19}$ $C$ per pulse (corresponding to
curve C in Fig.\ref{fig:chirp_Fi_4corr}) and the estimated pulse repetition
frequency of 82 MHz \cite{Zrenner02Nature} results in a small but measurable
value of about $10\times10^{-12}$ ampere. When the energy
transfer between the molecule and electron-phonon excitations
in the molecules and leads is present for the linear chirp, this control model can
complicate the Landau-Zerner transition to a decaying level, which were solved in
Refs.\cite{Fain} and \cite{Aku92}.
The relaxation parameters in the derived closed set of the equations of motion
do not depend on the exciting magnetic field and the phonon numbers.
The present theory could be supplied by an additional equation described by thermal bath, as was shown
 in Ref.\cite{Nit1}, in the case when the Rabi frequency $\Omega_{R}$ is much smaller than the bath correlation frequency
$\omega_{c}$. If molecular states $\varepsilon_{m}$ are far from the Fermi
levels of both leads, $\omega_{c}$ is determined by the frequency interval for
the system-bath interaction matrix elements $V_{km}^{(MK)}$ and $V_{kk^{\prime
}}^{(NK)}$ and the density of states of metal leads.  The approximation of
constant relaxation parameters, which do not depend on exciting
magnetic radiation, is consistent with the rotating wave approximation used in our theory.
The situation is different if we assume that the orbital energy molecular level is
pinned to the Fermi energy of a lead. This may lead to highly nonlinear
current voltage dependence \cite{Nitzan06JCP}. In this case, $\omega_{c}$ is
determined also by the frequency interval at which $f_{K}(\varepsilon)$ is
essentially changed by $\sim k_{B}T/\hbar$. $\Omega_{R}$ can be of the same order of
magnitude with $\omega_{c}$ in the rotating wave approximation, and the dependence of the relaxation
parameters on exciting magnetic field \cite{Schreiber06} has to be
included in the theory.

\section{conclusion}
\label{sec:Concl}
We have investigated a model process driven by magnetic fields in a
molecular system connecting metal leads.
We considered the general theoretical aspects of the interaction
of a molecular spin nanojunction with the magnetic
field within a single-electron model.
A simple conduction model for the HOMO and LUMO
of a molecule is given in the presence of the magnetic field
when the electronic distribution in
the molecule is far from equilibrium under bias voltage.
This nonequilibrium state is associated with the electron
flux between the source and leads, energy
flux between the nonequilibrium molecular electronic distribution,
and the local electronic distribution in
the metals leads. This is a nonradiative dissipation mechanism that
couples electronic excitations in the molecule to excitons in the metal.
We have investigated two aspects of the interrelationships
between electron fluxes. First, we have calculated the dependence
of the current pulse shape of the molecule on the parameters of Gaussian pulse.
 Second, we have studied the condition
under which current may be induced without bias voltage
by the external magnetic field to adjust a nonequilibrium steady state.
  Due to the competing relaxation processes, the
magnetic field induced current suggests
that the observation is feasible.
 The control of electron transfer by magnetic fields in a metal-molecule-metal spin nanojunction is
not a simple task. The theory can be useful for the development of the treatment
of negative differential resistance and giant magnetoresistance in molecular spin nanojunctions\cite{KimWY,Kim3,Kim2}.
The present results suggest that the experiments for an observation of ARP
are indeed feasible. We derived the explicit solution of general case for the current Eq.(\ref{eq:curr}).
It is shown that the system excited by magnetic field has an additional combination frequency.
The differential equations
 Eqs. (\ref{eq:n_sh}) and (\ref{eq:p_sh}) contain new
additional combinational frequencies depending on the concentrations
of molecular electrons and fermi functions. The additional combination part of the current
is the effective tool to measure the concentration of molecular electrons and it can be extended to
measuring the concentration of the charge carriers for another nanodevice.

    As was mentioned in Sec.\ref{sec:Introduction}, future generations of the spintronic
molecular systems would employ the coherent magnetic manipulations. We
considered such coherent control processes for the adjustment of $Q$ by pulse with independent parameters of
$\mu$ and $\delta$. Both the transfer driven current and charge that give rise to the
net current in the biased spin nanojunction and the energy transfer between the molecule
 and electron-phonon excitations in the molecules and leads can grow with growing $\mu$ and $\delta$ regardless of the loss of energy.
 It should be emphasized that all processes
considered in this work may play important roles in spin nanojunction in
response to the incident magnetic fields. First, direct electron-phonon excitations
of the molecular system and the metal leads \cite{Petek97} may affect the response to an
adsorbed molecule that is beyond the local field enhancement
associated with the local excitation. Second, experimental
realization of strong local excitations in nanojunctions requires
careful consideration of phonon dissipation and conduction
\cite{Nitzan07Jphys}. Phonon-electron interaction and phonon excitation
 may be kept under control by the junction using a sequence of well separated magnetic pulses, as
noted in the proposed experiment.

\textbf{Acknowledgement}
This work was supported by KOSEF (WCU: R32-2008-000-10180-0, EPB Center: R11-2008-052-01000), BK21(KRF), and GRL (KICOS).
We thank B. Fainberg for discussion.

\appendix
\section{Heisenberg equation for the electron - molecule interaction}
\label{sec:AppA}
We use the following Pauli commutators:
\begin{equation}
[\hat{c}_{i},\hat{c}_{j}^{+}]\equiv\hat{c}_{i}\hat{c}_{j}^{+}+\hat{c}%
_{j}^{+}\hat{c}_{i}=\delta_{ij}, \label{eq:anticommutation1}%
\end{equation}
\begin{equation}
[\hat{c}_{i},\hat{c}_{j}]\equiv[\hat{c}_{i}^{+},\hat{c}_{j}^{+}]\equiv
\hat{c}_{i}\hat{c}_{j}+\hat{c}_{j}\hat{c}_{i}=0.
\label{eq:anticommutation2}%
\end{equation}
where $\hat{c}_{i}$ and $\hat{c}^{+}_{j}$ are Fermi operators, $\delta_{ij}$ is the Kroenecker delta, and $\hat{c}_{i}^{+}\hat{c}%
_{i}=\hat{n}_{i}$, $\hat{c}_{i}\hat{c}_{i}^{+}=1-\hat{n}_{i}$.
For the unpertubed Hamiltonian $H_{0}$ in Eq.(\ref{eq:Ho}), we have
\begin{equation}
\lbrack\hat{H}_{0},\hat{c}_{m}]=-\sum_{m^{\prime}\in M}\varepsilon_{m^{\prime}}(\hat
{c}_{m^{\prime}}^{+}\hat{c}_{m}+\hat{c}_{m}\hat{c}_{m^{\prime}}^{+})\hat
{c}_{m^{\prime}}=-\varepsilon_{m}\hat{c}_{m}.
\end{equation}
The corresponding terms for $c_{k}(c_{k}^{+})$
in the Heisenberg equations is
\begin{equation}
\lbrack\hat{H}_{0},\hat{c}_{k}]=-\sum_{k^{\prime}\in K}\varepsilon_{k^{\prime}}(\hat
{c}_{k^{\prime}}^{+}\hat{c}_{k}+\hat{c}_{k}\hat{c}_{k^{\prime}}^{+})\hat
{c}_{k^{\prime}}=-\varepsilon_{k}\hat{c}_{k}.
\end{equation}
In Eq.(\ref{eq:Heis2}) the interaction term of electrons in the molecule $\hat{V}_{M}$ has
\begin{eqnarray}
\lbrack\hat{V}_{M},\hat{c}_{m}]=\sum_{K=L,R}%
\sum_{m,m^{\prime}\in M;k\in K}\{-V_{km^{\prime}}^{(MK)}(\hat{c}_{k}^{+}\hat
{c}_{m}+\hat{c}_{m}\hat{c}_{k}^{+})\hat{c}_{m^{\prime}}\nonumber\\
-V_{m^{\prime}k}^{(MK)}(\hat{c}_{m^{\prime}}^{+}\hat{c}_{m}+\hat{c}_{m}
\hat{c}_{m^{\prime}}^{+})\hat{c}_{k}\}
=-\sum_{K=L,R}\sum_{k\in K}V_{mk}^{(MK)}\hat
{c}_{k},\nonumber\\
\end{eqnarray}
and the interaction term of electrons in the leads has
\begin{eqnarray}
\lbrack\hat{V}_{M},\hat{c}_{k}]=\sum_{K=L,R}\sum_{m^{\prime
}\in M;k,k^{\prime}\in K}\{-V_{k^{\prime}m^{\prime}}^{(MK)}(\hat{c}_{k^{\prime}%
}^{+}\hat{c}_{k}+\hat{c}_{k}\hat{c}_{k^{\prime}}^{+})\hat{c}_{m^{\prime}}\nonumber\\
-V_{m^{\prime}k^{\prime}}^{(MK)}(\hat{c}_{m^{\prime}}^{+}\hat{c}_{k}
+\hat{c}_{k}\hat{c}_{m^{\prime}}^{+})\hat{c}_{k^{\prime}}\}\nonumber\\
=-\sum_{m^{\prime}\in M}V_{km^{\prime}}^{(MK)}\hat{c}_{m^{\prime}}.%
\end{eqnarray}
To simplify the representation of Hamiltonian Eq.(\ref{eq:V_M}) of the energy transfer,
we take into account $m\in M,$ where $M=\{1,2\}$ (where 1,2 denotes the HOMOs and LUMO).
Then, the Poisson brackets of
the electrons in the molecule have the form:
\begin{widetext}
\begin{eqnarray}
\lbrack\hat{V}_{N},\hat{c}_{m}]=\sum_{K=L,R}\sum_{k\neq
k^{\prime}\in K}\{V_{kk^{\prime}}^{(NK)}(\hat{c}_{k}^{+}\hat{c}_{k^{\prime}
}\hat{c}_{2}^{+}\hat{c}_{1}\hat{c}_{m}-\hat{c}_{m}\hat{c}_{k}^{+}\hat{c}_{k^{\prime}}\hat{c}_{2}^{+}\hat{c}
_{1})+V_{k^{\prime}k}^{(NK)}(\hat{c}_{1}^{+}\hat{c}_{2}\hat{c}_{k^{\prime}
}^{+}\hat{c}_{k}\hat{c}_{m}-\hat{c}_{m}\hat{c}_{1}^{+}\hat{c}_{2}\hat
{c}_{k^{\prime}}^{+}\hat{c}_{k})\}\nonumber\\
=-\sum_{K=L,R}\sum_{k\neq k^{\prime}\in K}\{V_{kk^{\prime}}^{(NK)}\hat
{c}_{k}^{+}\hat{c}_{k^{\prime}}(\hat{c}_{2}^{+}\hat{c}_{m}+\hat{c}_{m}\hat
{c}_{2}^{+})\hat{c}_{1}+V_{k^{\prime}k}^{(NK)}\delta_{1m}(\hat{c}_{1}^{+}
\hat{c}_{1}+\hat{c}_{1}\hat{c}_{1}^{+})\hat{c}_{2}\hat{c}_{k^{\prime}}^{+}
\hat{c}_{k}\}\nonumber\\
=-\sum_{K=L,R}\sum_{k\neq k^{\prime}\in K}\{V_{kk^{\prime}}^{(NK)}\hat
{c}_{k}^{+}\hat{c}_{k^{\prime}}\hat{c}_{1}\delta_{2m}+V_{k^{\prime}k}
^{(NK)}\delta_{1m}\hat{c}_{2}\hat{c}_{k^{\prime}}^{+}\hat{c}_{k}\}
\end{eqnarray}
\end{widetext}
The Poisson brackets for the lead electrons have the form:
\begin{widetext}
\begin{align*}
\lbrack\hat{V}_{N},\hat{c}_{k}]  &  =[\sum_{K=L,R}\sum_{k^{\prime\prime}\neq
k^{\prime}\in K}(V_{k^{\prime\prime}k^{\prime}}^{(NK)}\hat{c}_{k^{\prime
\prime}}^{+}\hat{c}_{k^{\prime}}\hat{c}_{2}^{+}\hat{c}_{1}+V_{k^{\prime
}k^{\prime\prime}}^{(NK)}\hat{c}_{1}^{+}\hat{c}_{2}\hat{c}_{k^{\prime}}%
^{+}\hat{c}_{k^{\prime\prime}}),\hat{c}_{k}]\\
&  =\sum_{K=L,R}\sum_{k^{\prime\prime}\neq k^{\prime}\in K}\{V_{k^{\prime
\prime}k^{\prime}}^{(NK)}(\hat{c}_{k^{\prime\prime}}^{+}\hat{c}_{k^{\prime}%
}\hat{c}_{2}^{+}\hat{c}_{1}\hat{c}_{k}-\hat{c}_{k}\hat{c}_{k^{\prime\prime}%
}^{+}\hat{c}_{k^{\prime}}\hat{c}_{2}^{+}\hat{c}_{1})+V_{k^{\prime}%
k^{\prime\prime}}^{(NK)}(\hat{c}_{1}^{+}\hat{c}_{2}\hat{c}_{k^{\prime}}%
^{+}\hat{c}_{k^{\prime\prime}}\hat{c}_{k}-\\
&  -\hat{c}_{k}\hat{c}_{1}^{+}\hat{c}_{2}\hat{c}_{k^{\prime}}^{+}\hat
{c}_{k^{\prime\prime}})\}\\
&  =\sum_{K=L,R}\sum_{k^{\prime\prime}\neq k^{\prime}\in K}\{V_{k^{\prime
\prime}k^{\prime}}^{(NK)}(\hat{c}_{k^{\prime\prime}}^{+}\hat{c}_{k^{\prime}%
}\hat{c}_{k}-\hat{c}_{k}\hat{c}_{k^{\prime\prime}}^{+}\hat{c}_{k^{\prime}%
})\hat{c}_{2}^{+}\hat{c}_{1}+V_{k^{\prime}k^{\prime\prime}}^{(NK)}\hat{c}%
_{1}^{+}\hat{c}_{2}(\hat{c}_{k^{\prime}}^{+}\hat{c}_{k^{\prime\prime}}\hat
{c}_{k}-\hat{c}_{k}\hat{c}_{k^{\prime}}^{+}\hat{c}_{k^{\prime\prime}})\}\nonumber\\
&-\{\sum_{k^{\prime}\neq k\in K}V_{kk^{\prime}}^{(NK)}\hat{c}_{k^{\prime}}\hat{c}_{2}^{+}\hat{c}_{1}%
+\sum_{k^{\prime\prime}\neq k\in K}V_{kk^{\prime\prime}}^{(NK)}\hat{c}_{1}%
^{+}\hat{c}_{2}\hat{c}_{k^{\prime\prime}}\}\\
&  =-\sum_{k^{\prime}\neq k\in K}V_{kk^{\prime}}^{(NK)}(\hat{c}_{k^{\prime}%
}\hat{c}_{2}^{+}\hat{c}_{1}+\hat{c}_{1}^{+}\hat{c}_{2}\hat{c}_{k^{\prime}%
})=-\sum_{k^{\prime}\neq k\in K}V_{kk^{\prime}}^{(NK)}(\hat{c}_{2}^{+}\hat
{c}_{1}+\hat{c}_{1}^{+}\hat{c}_{2})\hat{c}_{k^{\prime}}.%
\end{align*}
\end{widetext}
To obtain the formulas presented above, we used the following rules of the triadic
multiplication for operators.
If $k\neq k^{\prime}$, then $(\hat{c}_{k^{\prime\prime}}^{+}\hat{c}%
_{k^{\prime}}\hat{c}_{k}-\hat{c}_{k}\hat{c}_{k^{\prime\prime}}^{+}\hat
{c}_{k^{\prime}})=-(\hat{c}_{k^{\prime\prime}}^{+}\hat{c}_{k}+\hat{c}_{k}%
\hat{c}_{k^{\prime\prime}}^{+})\hat{c}_{k^{\prime}}=-\delta_{kk^{\prime\prime
}}\hat{c}_{k^{\prime}}$.
If $k=k^{\prime}$, then $(\hat{c}_{k^{\prime\prime}}^{+}\hat{c}_{k}\hat{c}%
_{k}-\hat{c}_{k}\hat{c}_{k^{\prime\prime}}^{+}\hat{c}_{k})=-\hat{c}_{k}\hat
{c}_{k^{\prime\prime}}^{+}\hat{c}_{k}=\hat{c}_{k^{\prime\prime}}^{+}\hat
{c}_{k}\hat{c}_{k}=0$, since $k''\neq k'$.
Therefore, we have
$(\hat{c}_{k^{\prime\prime}}^{+}\hat{c}_{k^{\prime}}\hat{c}_{k}-\hat{c}_{k}%
\hat{c}_{k^{\prime\prime}}^{+}\hat{c}_{k^{\prime}})=-\delta_{kk^{\prime\prime
}}\delta_{k^{\prime}k^{\prime\prime}}\hat{c}_{k^{\prime}}$
for $k^{\prime\prime}\neq k^{\prime}$.
If $k\neq k^{\prime\prime}$, then $(\hat{c}_{k^{\prime}}^{+}\hat{c}%
_{k^{\prime\prime}}\hat{c}_{k}-\hat{c}_{k}\hat{c}_{k^{\prime}}^{+}\hat
{c}_{k^{\prime\prime}})=-(\hat{c}_{k^{\prime}}^{+}\hat{c}_{k}+\hat{c}_{k}%
\hat{c}_{k^{\prime}}^{+})\hat{c}_{k^{\prime\prime}}=-\delta_{kk^{\prime}}%
\hat{c}_{k^{\prime\prime}}$.
If $k=k^{\prime\prime}$, then $(\hat{c}_{k^{\prime}}^{+}\hat{c}_{k}\hat{c}%
_{k}-\hat{c}_{k}\hat{c}_{k^{\prime}}^{+}\hat{c}_{k})=-\hat{c}_{k}\hat
{c}_{k^{\prime}}^{+}\hat{c}_{k}=0$.
Therefore, we have
$(\hat{c}_{k^{\prime}}^{+}\hat{c}_{k^{\prime\prime}}\hat{c}_{k}-\hat{c}_{k}%
\hat{c}_{k^{\prime}}^{+}\hat{c}_{k^{\prime\prime}})=-\delta_{kk^{\prime}}\delta_{kk^{\prime\prime}}
\hat{c}_{k^{\prime\prime}}$.

\section{Heisenberg equation for the electron - magnetic field interaction}
\label{sec:AppB}
In the Heisenberg equation with the magnetic fields (\ref{eq:Heis2}), the term of $\hat{H}_{B}$ in Eq.(\ref{eq:hb1}) for the interaction of
the electron in the molecular system $c_{m}$
has the form:
\begin{eqnarray}
\hat{H}_{B}=-\mathbf{H}_{B}(\mathbf{r},t)(\hat{c}_{2}^{+}\hat
{c}_{1}+\hat{c}_{1}^{+}\hat{c}_{2}\mathcal{)}\nonumber\\
=-\frac{1}{2}(g\mathbf{\mu}_{B}
\cdot\mathbf{e}\{\hat{c}_{2}^{+}\hat{c}_{1}\mathcal{B}(t)\exp[-i\omega
t+i\varphi(t)]\nonumber\\
+\hat{c}_{1}^{+}\hat{c}_{2}\mathcal{B}^{\ast}(t)\exp[i\omega
t-i\varphi(t)]\}
\label{eq:hbField}
\end{eqnarray}
where $\mathbf{e}$ is the unit vector. Substituting Eq.(\ref{eq:hbField}) in the Poisson brackets of Eq.(\ref{eq:Heis2}), we have
\begin{widetext}
\begin{eqnarray}
\lbrack\hat{H}_{B},\hat{c}_{m}]=-\frac{1}{2}(g\mathbf{\mu}_{B}\cdot
\mathbf{e)}[\hat{c}_{2}^{+}\hat{c}_{1}\mathcal{B}(t)\exp[-i\omega
t+i\varphi(t)]+\hat{c}_{1}^{+}\hat{c}_{2}\mathcal{B}^{\ast}(t)\exp[i\omega
t-i\varphi(t)],\hat{c}_{m}]\nonumber\\
=-\frac{1}{2}(g\mathbf{\mu}_{B}\cdot\mathbf{e)}\{(\hat{c}_{2}^{+}\hat{c}_{1}%
\hat{c}_{m}-\hat{c}_{m}\hat{c}_{2}^{+}\hat{c}_{1})\mathcal{B}(t)\exp[-i\omega
t+i\varphi(t)]+\nonumber\\
(\hat{c}_{1}^{+}\hat{c}_{2}\hat{c}_{m}-\hat{c}_{m}\hat{c}%
_{1}^{+}\hat{c}_{2})\mathcal{B}^{\ast}(t)\exp[i\omega t-i\varphi(t)]\}.
\label{eq:HbField2}
\end{eqnarray}
\end{widetext}
Taking into account that $m\in M, M=\{1,2\}$ for Eq.(\ref{eq:HbField2}), we have:
\begin{align*}
\lbrack\hat{H}_{B},\hat{c}_{1}]
=\frac{1}{2}(g\mathbf{\mu}_{B}\cdot\mathbf{e)}\mathcal{B}^{\ast}(t)\exp[i\omega
t-i\varphi(t)]\hat{c}_{2}%
\end{align*}
and
\begin{align*}
\lbrack\hat{H}_{B},\hat{c}_{2}]
=\frac{1}{2}(g\mathbf{\mu}_{B}\cdot\mathbf{e)}\mathcal{B}(t)\exp[-i\omega
t+i\varphi(t)]\hat{c}_{1}.%
\end{align*}
Similar to this, the Poisson brackets of Eq.(\ref{eq:Heis2})
for the magnetic field and the lead electrons yield:
\begin{align*}
\lbrack\hat{H}_{B},\hat{c}_{k}]=-\mathbf{H}_{B}(\mathbf{r}%
,t)[\hat{c}_{2}^{+}\hat{c}_{1}+\hat{c}_{1}^{+}\hat{c}_{2},\hat{c}_{k}]\\
=-\mathbf{H}_{B}(\mathbf{r},t)\{(\hat{c}_{2}^{+}\hat{c}_{1}%
\hat{c}_{k}-\hat{c}_{k}\hat{c}_{2}^{+}\hat{c}_{1})+(\hat{c}_{1}^{+}\hat{c}%
_{2}\hat{c}_{k}-\hat{c}_{k}\hat{c}_{1}^{+}\hat{c}_{2})\}\\
=-\mathbf{H}_{B}(\mathbf{r},t)\{(\hat{c}_{2}^{+}\hat{c}_{1}%
-\hat{c}_{2}^{+}\hat{c}_{1})\hat{c}_{k}+(\hat{c}_{1}^{+}\hat{c}_{2}-\hat
{c}_{1}^{+}\hat{c}_{2})\hat{c}_{k}\}=0.
\end{align*}
\section{Heisenberg equation for the electron - phonon interactions in the molecule}
\label{sec:AppC}
In the Heisenberg equation (\ref{eq:Heis2}), the term of $\hat{V}_{V}$ in the interaction of
molecular phonons $\hat{Q}_{\alpha}^{a}$ and electrons $c_{k}$ has the form:
\begin{equation}
\frac{d\hat{c}_{m}}{dt}=\frac{i}{\hbar}\left[\hat{V}_{V},\hat{c}_{m}\right]=
-\frac{i}{\hbar}\sum_{m\neq m^{\prime}\in M}(V_{mm^{\prime}}
^{(VK)}\hat{Q}_{\alpha}^{a})\hat{c}_{m},
\end{equation}
\begin{equation}
\frac{d\hat{c}_{m}^{+}}{dt}=\frac{i}{\hbar}\left[\hat{V}_{V},\hat{c}_{m}^{+}\right]=
-\frac{i}{\hbar}\sum_{m\neq m^{\prime}\in M}(V_{mm^{\prime}}
^{(VK)}\hat{Q}_{\alpha}^{a})\hat{c}_{m}^{+}.
\end{equation}
Following the phonon commutation rules $[a_{\alpha},a_{\alpha}^{+}]=1$ for the Poisson brackets of
the interaction of molecular phonons and electrons, we have
\begin{equation}
\left[\hat{H}_{0},\hat{c}_{k}\hat{Q}_{\alpha}^{a}\hat{c}_{m}^{+}\right]=
-\sum_{\alpha}i\omega_{\alpha}\hat{c}_{k}\hat{P}_{\alpha}^{a}\hat{c}_{m}^{+},
\label{eq:cQc+}
\end{equation}
\begin{eqnarray}
\left[\hat{H}_{0},\hat{c}_{k}^{+}\hat{Q}_{\alpha}^{a}\hat{c}_{m}\right]=
-\sum_{\alpha}i\omega_{\alpha}\hat{c}_{k}^{+}\hat{P}_{\alpha}^{a}\hat{c}_{m}\nonumber\\
+\sum_{m\in M, k\in K}(\varepsilon_{k}-\varepsilon_{m})\hat{c}_{k}^{+}\hat{Q}_{\alpha}^{a}\hat{c}_{m}\nonumber\\
+\sum_{m\in M, k\in K}2(\varepsilon_{m}n_{m}-\varepsilon_{k}f_{K}(\varepsilon_{k}))
\hat{c}_{k}^{+}\hat{Q}_{\alpha}^{a}\hat{c}_{m},
\label{eq:c+Qc}
\end{eqnarray}
\begin{equation}
\left[\hat{H}_{0},\hat{c}_{k}\hat{P}_{\alpha}^{a}\hat{c}_{m}^{+}\right]=
\sum_{\alpha}i\omega_{\alpha}\hat{c}_{k}\hat{Q}_{\alpha}^{a}\hat{c}_{m}^{+},
\end{equation}
\begin{eqnarray}
\left[\hat{H}_{0},\hat{c}_{k}^{+}\hat{P}_{\alpha}^{a}\hat{c}_{m}\right]=
\sum_{\alpha}i\omega_{\alpha}\hat{c}_{k}^{+}\hat{Q}_{\alpha}^{a}\hat{c}_{m}\nonumber\\
+\sum_{m\in M, k\in K}(\varepsilon_{k}-\varepsilon_{m})\hat{c}_{k}^{+}\hat{P}_{\alpha}^{a}\hat{c}_{m}\nonumber\\
+\sum_{m\in M, k\in K}2(\varepsilon_{m}n_{m}-\varepsilon_{k}f_{K}(\varepsilon_{k}))
\hat{c}_{k}^{+}\hat{P}_{\alpha}^{a}\hat{c}_{m},
\end{eqnarray}
\begin{eqnarray}
\left[\hat{V}_{M},\hat{c}_{k}\hat{Q}_{\alpha}^{a}\hat{c}_{m}^{+}\right]=%
2\sum_{K=L,R}\sum_{m\in M,k\in K}V_{km}^{(MK)}
((1-n_{m})f_{K}(\varepsilon_{k})\nonumber\\
-(1-f_{K}(\varepsilon_{k}))n_{m})Q_{\alpha},%
\label{eq:cQc+M}
\end{eqnarray}
\begin{eqnarray}
\left[\hat{V}_{M},\hat{c}_{k}\hat{P}_{\alpha}^{a}\hat{c}_{m}^{+}\right]=%
2\sum_{K=L,R}\sum_{m\in M,k\in K}V_{km}^{(MK)}
((1-n_{m})f_{K}(\varepsilon_{k})\nonumber\\
-(1-f_{K}(\varepsilon_{k}))n_{m})P_{\alpha},%
\end{eqnarray}
\begin{eqnarray}
\left[\hat{V}_{M},\hat{c}_{k}^{+}\hat{Q}_{\alpha}^{a}\hat{c}_{m}\right]=%
\left[\hat{V}_{M},\hat{c}_{k}^{+}\hat{P}_{\alpha}^{a}\hat{c}_{m}\right]=0%
\end{eqnarray}
\begin{equation}
\left[\hat{V}_{V},\hat{c}_{k}\hat{Q}_{\alpha}^{a}\hat{c}_{m}^{+}\right]=
\sum_{m\neq m'\in M}V_{mm'}^{(VK)}(1-2n_{m})
\hat{c}_{k}\hat{c}_{m}^{+}Q_{\alpha}^{2},
\label{eq:cQc+E}
\end{equation}
\begin{equation}
\left[\hat{V}_{V},\hat{c}_{k}^{+}\hat{Q}_{\alpha}^{a}\hat{c}_{m}\right]=
\sum_{m\neq m'\in M}V_{mm'}^{(VK)}(2n_{m}+1)\hat{c}_{k}^{+}\hat{c}_{m}Q_{\alpha}^{2},
\end{equation}
\begin{eqnarray}
\left[\hat{V}_{V},\hat{c}_{k}\hat{P}_{\alpha}^{a}\hat{c}_{m}^{+}\right]=
\left[\hat{V}_{V},\hat{c}_{k}^{+}\hat{P}_{\alpha}^{a}\hat{c}_{m}\right]=\nonumber\\
2i\sum_{m\neq m'\in M}V_{mm'}^{(VK)}
\hat{c}_{k}^{+}\hat{c}_{m},
\end{eqnarray}

\section{Calculation of $\langle\hat{c}_{k}^{+}\hat{c}_{m}\rangle$
in the presence of a magnetic field}
\label{sec:AppD}
In this section we calculate the expression for the expectation value
$\langle\hat{c}_{k}^{+}\hat{c}_{m}\rangle$
by taking into account the magnetic field,
namely using Eqs. (\ref{eq:cm}), (\ref{eq:cpm}),
(\ref{eq:cm+dcm}) and (\ref{eq:dcm+cm}).
For $m=1$, we obtain
\begin{widetext}
\begin{align}
\frac{d}{dt}\langle\hat{c}_{k}^{+}\hat{c}_{1}\rangle &  =\frac{i}{\hbar
}(\varepsilon_{k}-\varepsilon_{1})\langle\hat{c}_{k}^{+}\hat{c}_{1}%
\rangle+\frac{i}{\hbar}\sum_{m^{\prime}=1,2}V_{m^{\prime}k}^{(MK)}\langle
\hat{c}_{m^{\prime}}^{+}\hat{c}_{1}\rangle-\frac{i}{\hbar}\sum_{K=L,R}%
\sum_{k^{\prime}\in K}V_{1k^{\prime}}^{(MK)}\langle\hat{c}_{k}^{+}\hat
{c}_{k^{\prime}}\rangle\nonumber\\
&  +\frac{i}{\hbar}\sum_{k^{\prime}\neq k\in K}V_{k^{\prime}k}^{(NK)}%
\langle\hat{c}_{k^{\prime}}^{+}\hat{c}_{1}^{+}\hat{c}_{2}\hat{c}_{1}%
\rangle+\frac{i}{2\hbar}(g\mathbf{\mu}_{B}\cdot\mathbf{e)}\mathcal{B}^{\ast}%
(t)\exp[i\omega t-i\varphi(t)]\langle\hat{c}_{k}^{+}\hat{c}_{2}\rangle
\label{eq:dc^+_kc_1}\\
&  -\frac{i}{\hbar}\sum_{K=L,R}\sum_{k^{\prime\prime}\neq k^{\prime}\in
K}\langle V_{k^{\prime}k^{\prime\prime}}^{(NK)}\hat{c}_{k}^{+}\hat{c}_{2}%
\hat{c}_{k^{\prime}}^{+}\hat{c}_{k^{\prime\prime}}\rangle.\nonumber
\end{align}
\end{widetext}
Since the term $\langle\hat{c}_{m^{\prime}}^{+}\hat{c}_{m}\rangle|_{m\neq m^{\prime}}$ is the
non-resonant case for multiplication of $\varepsilon_{k}-\varepsilon_{2}$ and $\varepsilon
_{k}-\varepsilon_{1}$, we must put $\langle\hat{c}_{m^{\prime}}%
^{+}\hat{c}_{m}\rangle=\langle\hat{c}_{m}^{+}\hat{c}_{m'}\rangle\delta
_{mm^{\prime}}$ in the second term on the right-hand side of
Eqs. (\ref{eq:dc^+_kc_1}) and (\ref{eq:dc^+_kc_2}).
For $m=2$,
\begin{widetext}
\begin{eqnarray}
\frac{d}{dt}\langle\hat{c}_{k}^{+}\hat{c}_{2}\rangle=\frac{i}{\hbar
}(\varepsilon_{k}-\varepsilon_{2})\langle\hat{c}_{k}^{+}\hat{c}_{2}
\rangle+\frac{i}{\hbar}\sum_{m^{\prime}=1,2}V_{m^{\prime}k}^{(MK)}\langle
\hat{c}_{m^{\prime}}^{+}\hat{c}_{2}\rangle\nonumber\\
-\frac{i}{\hbar}\sum_{K=L,R}
\sum_{k^{\prime}\in K}V_{2k^{\prime}}^{(MK)}\langle\hat{c}_{k}^{+}\hat
{c}_{k^{\prime}}\rangle\nonumber\\
+\frac{i}{\hbar}\sum_{k^{\prime}\neq k\in K}V_{k^{\prime}k}^{(NK)}
\langle\hat{c}_{k^{\prime}}^{+}\hat{c}_{2}^{+}\hat{c}_{1}\hat{c}_{2}\rangle\nonumber\\
+\frac{i}{2\hbar}(g\mathbf{\mu}_{B}\cdot\mathbf{e)}\mathcal{B}(t)\exp[-i\omega
t+i\varphi(t)]\langle\hat{c}_{k}^{+}\hat{c}_{1}\rangle-\frac{i}{\hbar}
\sum_{K=L,R}\sum_{k^{\prime\prime}\neq k^{\prime}\in K}\langle V_{k^{\prime
\prime}k^{\prime}}^{(NK)}\hat{c}_{k}^{+}\hat{c}_{k^{\prime\prime}}^{+}\hat
{c}_{k^{\prime}}\hat{c}_{1}\rangle\nonumber\\
\label{eq:dc^+_kc_2}
\end{eqnarray}
\end{widetext}
In addition, using Eq.(\ref{eq:c^+_kc_k'}) and
disregarding terms of $\sim V^{(NK)}$ as a first step, we obtain%
\begin{widetext}
\begin{eqnarray}
\frac{d}{dt}\langle\hat{c}_{k}^{+}\hat{c}_{m}\rangle &  =\frac{i}{\hbar
}(\varepsilon_{k}-\varepsilon_{m})\langle\hat{c}_{k}^{+}\hat{c}_{m}%
\rangle+\frac{i}{\hbar}V_{mk}^{(MK)}n_{m}-\frac{i}{\hbar}V_{mk}^{(MK)}%
f_{K}(\varepsilon_{k})\nonumber\\
+\frac{i}{2\hbar}(g\mathbf{\mu}_{B}\cdot\mathbf{e)}\mathcal{B}^{\ast}%
(t)\exp[i\omega t-i\varphi(t)]\langle\hat{c}_{k}^{+}\hat{c}_{m'\neq m}\rangle
\label{eq:dcen}
\end{eqnarray}
\end{widetext}
%
In order to derive the solution of Eqs.(\ref{eq:dc^+_kc_1},\ref{eq:dc^+_kc_2})
we shall use the slowly varying amplitude method. Lets us define
\begin{align}
\langle\hat{c}_{k}^{+}\hat{c}_{m}\rangle=\langle\hat{c}_{k}^{+}\hat{c}%
_{m}\rangle_{0}\exp[i\omega t-i\varphi(t)],
\label{eq:slvaram}
\end{align}%
by substituting Eq.(\ref{eq:slvaram}) for Eq.(\ref{eq:dcen}), we have
\begin{widetext}
\begin{align}
\frac{d}{dt}\langle\hat{c}_{k}^{+}\hat{c}_{m}\rangle=
\frac{i}{\hbar}(\varepsilon_{k}-\varepsilon_{1})\langle\hat{c}_{k}^{+}%
\hat{c}_{m}\rangle_{0}\exp[i\omega t-i\varphi(t)]+\frac{i}{\hbar}%
V_{mk}^{(MK)}n_{m}-\frac{i}{\hbar}V_{mk}^{(MK)}f_{K}(\varepsilon
_{k})+\nonumber\\
+\frac{i}{2\hbar}(g\mathbf{\mu}_{B}\cdot\mathbf{e)}\mathcal{B}^{\ast}%
(t)\exp[i\omega t-i\varphi(t)]\langle\hat{c}_{k}^{+}\hat{c}_{m'\neq m}\rangle.
\label{eq:cmk}
\end{align}
\end{widetext}
Conserving only the resonance terms in Eq.(\ref{eq:cmk}), we have%
\begin{widetext}
\begin{equation}
\frac{d}{dt}\langle\hat{c}_{k}^{+}\hat{c}_{m}\rangle_{0}=i[(\varepsilon
_{k}-\varepsilon_{m})/\hbar-\omega(t)]\langle\hat{c}_{k}^{+}\hat{c}_{m}%
\rangle_{0}+\frac{i}{2\hbar}(g\mathbf{\mu}_{B}\cdot\mathbf{e)}\mathcal{B}^{\ast
}(t)\langle\hat{c}_{k}^{+}\hat{c}_{m'\neq m}\rangle\label{eq:dc^+_kc_1int}.
\end{equation}
\end{widetext}
Integrating the last equation, we obtain%
\begin{widetext}
\begin{equation}
\langle\hat{c}_{k}^{+}\hat{c}_{m}\rangle_{0}=\frac{i}{2\hbar}(g\mathbf{\mu}_{B}%
\cdot\mathbf{e)}\mathcal{B}^{\ast}(t)\langle\hat{c}_{k}^{+}\hat{c}_{m'\neq m}%
\rangle\int_{0}^{\infty}d\tau\exp[i((\varepsilon_{k}-\varepsilon_{m}%
)/\hbar-\omega(t))\tau],
\end{equation}
\end{widetext}
\begin{widetext}
\begin{align*}
\frac{d}{dt}\langle\hat{c}_{k}^{+}\hat{c}_{m}\rangle=\frac{i}{\hbar
}(\varepsilon_{k}-\varepsilon_{m})\langle\hat{c}_{k}^{+}\hat{c}_{m}%
\rangle+\frac{i}{\hbar}V_{mk}^{(MK)}n_{m}-\frac{i}{\hbar}V_{mk}^{(MK)}%
f_{K}(\varepsilon_{k})+\frac{i}{2\hbar}(g\mathbf{\mu}_{B}\cdot\mathbf{e)}%
\mathcal{B}(t)\langle\hat{c}_{k}^{+}\hat{c}_{m'\neq m}\rangle_{0},
\label{eq:dc^+_kc_2a}%
\end{align*}
\end{widetext}
and
\begin{widetext}
\begin{align}
\langle\hat{c}_{k}^{+}\hat{c}_{m}\rangle &  \simeq\frac{i}{\hbar}V_{mk}%
^{(MK)}[n_{m}(t)-f_{K}(\varepsilon_{k})]\int_{0}^{\infty}d\tau\exp[\frac
{i}{\hbar}(\varepsilon_{k}-\varepsilon_{m})\tau]-\\
&  -\frac{1}{4\hbar^{2}}(g\mathbf{\mu}_{B}\cdot\mathbf{e)}^{2}|\mathcal{B}%
(t)|^{2}\langle\hat{c}_{k}^{+}\hat{c}_{m}\rangle\int_{0}^{\infty}d\tau
\exp[\frac{i}{\hbar}(\varepsilon_{k}-\varepsilon_{m})\tau]\int_{0}^{\infty
}d\tau\exp[i((\varepsilon_{k}-\varepsilon_{m'\neq m})/\hbar-\omega(t))\tau].
\end{align}
\end{widetext}
or%
\begin{widetext}
\begin{align}
\langle\hat{c}_{k}^{+}\hat{c}_{m}\rangle\
\cong\frac{i}{\hbar}V_{mk}^{(MK)}[n_{m}(t)-f_{K}(\varepsilon_{k})]\int
_{0}^{\infty}d\tau\exp[\frac{i}{\hbar}(\varepsilon_{k}-\varepsilon_{m})\tau]\nonumber\\
\{{1-\frac{1}{4\hbar^{2}}}(g\mathbf{\mu}_{B}\cdot\mathbf{e)}^{2}|\mathcal{B}(t)|^{2}\int_{0}^{\infty}d\tau
\exp[\frac{i}{\hbar}(\varepsilon_{k}-\varepsilon_{m})\tau]\int_{0}^{\infty
}d\tau\exp[i((\varepsilon_{k}-\varepsilon_{m'\neq m})/\hbar-\omega(t))\tau]\}
\label{eq:aprck}
\end{align}
\end{widetext}
The last equation gives "saturation" corrections to the preliminary result
Eq.(\ref{eq:c^+_kc_m_1}) by taking into account the external magnetic fields.
The approximation for the magnetic field dependent series could be produced by formal integration of Eq.(\ref{eq:aprck})
with an appropriate treatment related to the magnetic field dependent relaxation parameters, namely
\begin{equation}
\langle\hat{c}_{k}^{+}\hat{c}_{m}\rangle\cong(n_{m}(t)-f_{K}(\varepsilon_{k}))
(1-|\mathcal{B}(t)|^{2}\Gamma_{BK,m})\Gamma_{MK,m},
\end{equation}
where
\begin{equation}
\Gamma_{BK,m}=\frac{\hbar^{2}(g\mathbf{\mu}_{B}\cdot\mathbf{e)}^{2}P^{2}}{(\varepsilon_{k}-\varepsilon_{m})^{2}}\Gamma_{MK,m}.
\end{equation}

\section{Simplified equations of current for the $\hat{V}_{M}$ interaction}
\label{sec:AppC}
In this section we derive the simplified formula for the current and set
the differential equations for $n_{1}$ and $n_{2}$ without $\hat{V}_{V}$,
$\hat{V}_{N}$ and $\hat{H}_{B}$. Using the formula for the current Eq.(\ref{eq:curr}),
we obtain
\begin{align}
I  &  =e\{[-f_{L}(\varepsilon_{1})+n_{1}]\Gamma_{ML,1}+[n_{2}%
-f_{L}(\varepsilon_{2})]\Gamma_{ML,2}\}.
\end{align}
For $K\in{L,R}$, the damping coefficient has the form:
\begin{equation}
\Gamma_{MK,1}=\frac{2\pi}{\hbar}\sum_{k\in K}|V_{1k}^{(MK)}|^{2}%
\delta(\varepsilon_{1k}-\hbar\omega_{0}/2),
\end{equation}
\begin{equation}
\Gamma_{MK,2}=\frac{2\pi}{\hbar}\sum_{k\in K}|V_{2,k}^{(MK)}|^{2}\delta
(\varepsilon_{2k}-\hbar\omega_{0}/2).
\end{equation}
The ordinary differential equation Eq.(\ref{eq:n_eM2}) has the simplified forms:
\begin{equation}
\frac{dn_{1}}{dt}=\Omega_{R}\operatorname{Im}p_{M}+\sum_{K=L,R}%
[f_{1K}(\hbar\omega_{0}/2)-n_{1}]\Gamma_{MK,1}%
\end{equation}
\begin{equation}
\frac{dn_{2}}{dt}=\Omega_{R}\operatorname{Im}p_{M}-\sum_{K=L,R}%
[n_{2}-f_{2K}(\hbar\omega_{0}/2)]\Gamma_{MK,2}.%
\end{equation}
The differential equation for the polarization Eq.(\ref{eq:dpm}) has the form:
\begin{widetext}
\begin{equation}
\frac{dp_{M}}{dt}=-i[\omega_{0}+\sum_{K=L,R}\Delta_{MK}-\omega
(t)]p_{M}-\frac{i}{2}\Omega_{R}(n_{1}+n_{2}-1)-\frac{1}{2}\sum
_{K=L,R}(\Gamma_{MK,2}+\Gamma_{MK,1})p_{M}.
\end{equation}
\end{widetext}


\begin{thebibliography}{78}
\expandafter\ifx\csname natexlab\endcsname\relax\def\natexlab#1{#1}\fi
\expandafter\ifx\csname bibnamefont\endcsname\relax
  \def\bibnamefont#1{#1}\fi
\expandafter\ifx\csname bibfnamefont\endcsname\relax
  \def\bibfnamefont#1{#1}\fi
\expandafter\ifx\csname citenamefont\endcsname\relax
  \def\citenamefont#1{#1}\fi
\expandafter\ifx\csname url\endcsname\relax
  \def\url#1{\texttt{#1}}\fi
\expandafter\ifx\csname urlprefix\endcsname\relax\def\urlprefix{URL }\fi
\providecommand{\bibinfo}[2]{#2}
\providecommand{\eprint}[2][]{\url{#2}}

\bibitem[{\citenamefont{Binasch et~al.}(1989)\citenamefont{Binasch, Gr\"unberg,
  Saurenbach, and Zinn}}]{Binasch}
\bibinfo{author}{\bibfnamefont{G.}~\bibnamefont{Binasch}},
  \bibinfo{author}{\bibfnamefont{P.}~\bibnamefont{Gr\"unberg}},
  \bibinfo{author}{\bibfnamefont{F.}~\bibnamefont{Saurenbach}},
  \bibnamefont{and} \bibinfo{author}{\bibfnamefont{W.}~\bibnamefont{Zinn}},
  \bibinfo{journal}{Phys. Rev. B} \textbf{\bibinfo{volume}{39}},
  \bibinfo{pages}{4828} (\bibinfo{year}{1989}).

\bibitem[{\citenamefont{Baibich et~al.}(1988)\citenamefont{Baibich, Broto,
  Fert, Van~Dau, Petroff, Etienne, Creuzet, Friederich, and
  Chazelas}}]{Baibich}
\bibinfo{author}{\bibfnamefont{M.~N.} \bibnamefont{Baibich}},
  \bibinfo{author}{\bibfnamefont{J.~M.} \bibnamefont{Broto}},
  \bibinfo{author}{\bibfnamefont{A.}~\bibnamefont{Fert}},
  \bibinfo{author}{\bibfnamefont{F.~N.} \bibnamefont{Van~Dau}},
  \bibinfo{author}{\bibfnamefont{F.}~\bibnamefont{Petroff}},
  \bibinfo{author}{\bibfnamefont{P.}~\bibnamefont{Etienne}},
  \bibinfo{author}{\bibfnamefont{G.}~\bibnamefont{Creuzet}},
  \bibinfo{author}{\bibfnamefont{A.}~\bibnamefont{Friederich}},
  \bibnamefont{and} \bibinfo{author}{\bibfnamefont{J.}~\bibnamefont{Chazelas}},
  \bibinfo{journal}{Phys. Rev. Lett.} \textbf{\bibinfo{volume}{61}},
  \bibinfo{pages}{2472} (\bibinfo{year}{1988}).

\bibitem[{\citenamefont{Parkin et~al.}(1990)\citenamefont{Parkin, More, and
  Roche}}]{Parkin}
\bibinfo{author}{\bibfnamefont{S.~S.~P.} \bibnamefont{Parkin}},
  \bibinfo{author}{\bibfnamefont{N.}~\bibnamefont{More}}, \bibnamefont{and}
  \bibinfo{author}{\bibfnamefont{K.~P.} \bibnamefont{Roche}},
  \bibinfo{journal}{Phys. Rev. Lett.} \textbf{\bibinfo{volume}{64}},
  \bibinfo{pages}{2304} (\bibinfo{year}{1990}).

\bibitem[{\citenamefont{Asano et~al.}(1993)\citenamefont{Asano, Oguri, and
  Maekawa}}]{Asano}
\bibinfo{author}{\bibfnamefont{Y.}~\bibnamefont{Asano}},
  \bibinfo{author}{\bibfnamefont{A.}~\bibnamefont{Oguri}}, \bibnamefont{and}
  \bibinfo{author}{\bibfnamefont{S.}~\bibnamefont{Maekawa}},
  \bibinfo{journal}{Phys. Rev. B} \textbf{\bibinfo{volume}{48}},
  \bibinfo{pages}{6192} (\bibinfo{year}{1993}).

\bibitem[{\citenamefont{Moodera et~al.}(1995)\citenamefont{Moodera, Kinder,
  Wong, and Meservey}}]{Moodera1}
\bibinfo{author}{\bibfnamefont{J.~S.} \bibnamefont{Moodera}},
  \bibinfo{author}{\bibfnamefont{L.~R.} \bibnamefont{Kinder}},
  \bibinfo{author}{\bibfnamefont{T.~M.} \bibnamefont{Wong}}, \bibnamefont{and}
  \bibinfo{author}{\bibfnamefont{R.}~\bibnamefont{Meservey}},
  \bibinfo{journal}{Phys. Rev. Lett.} \textbf{\bibinfo{volume}{74}},
  \bibinfo{pages}{3273} (\bibinfo{year}{1995}).

\bibitem[{\citenamefont{Moodera et~al.}(1998)\citenamefont{Moodera, Nowak, and
  van~de Veerdonk}}]{Moodera2}
\bibinfo{author}{\bibfnamefont{J.~S.} \bibnamefont{Moodera}},
  \bibinfo{author}{\bibfnamefont{J.}~\bibnamefont{Nowak}}, \bibnamefont{and}
  \bibinfo{author}{\bibfnamefont{R.~J.~M.} \bibnamefont{van~de Veerdonk}},
  \bibinfo{journal}{Phys. Rev. Lett.} \textbf{\bibinfo{volume}{80}},
  \bibinfo{pages}{2941} (\bibinfo{year}{1998}).

\bibitem[{\citenamefont{Mehrez et~al.}(2000)\citenamefont{Mehrez, Taylor, Guo,
  Wang, and Roland}}]{Mehrez}
\bibinfo{author}{\bibfnamefont{H.}~\bibnamefont{Mehrez}},
  \bibinfo{author}{\bibfnamefont{J.}~\bibnamefont{Taylor}},
  \bibinfo{author}{\bibfnamefont{H.}~\bibnamefont{Guo}},
  \bibinfo{author}{\bibfnamefont{J.}~\bibnamefont{Wang}}, \bibnamefont{and}
  \bibinfo{author}{\bibfnamefont{C.}~\bibnamefont{Roland}},
  \bibinfo{journal}{Phys. Rev. Lett.} \textbf{\bibinfo{volume}{84}},
  \bibinfo{pages}{2682} (\bibinfo{year}{2000}).

\bibitem[{\citenamefont{Kim and Kim}(2008{\natexlab{a}})}]{KimWY}
\bibinfo{author}{\bibfnamefont{W.}~\bibnamefont{Kim}} \bibnamefont{and}
  \bibinfo{author}{\bibfnamefont{K.}~\bibnamefont{Kim}},
  \bibinfo{journal}{Nature Nanotech.} \textbf{\bibinfo{volume}{3}},
  \bibinfo{pages}{408} (\bibinfo{year}{2008}{\natexlab{a}}).

\bibitem[{\citenamefont{Gamov}(1928)}]{Gam}
\bibinfo{author}{\bibfnamefont{G.}~\bibnamefont{Gamov}}, \bibinfo{journal}{Z.
  Phys} \textbf{\bibinfo{volume}{51}}, \bibinfo{pages}{204}
  (\bibinfo{year}{1928}).

\bibitem[{Kag(2004)}]{Kagan_Ratner04}
\bibinfo{journal}{MRS Bull. Special issue on molecular junctions, edited by C.
  R. Kagan and M. A. Ratner.} \textbf{\bibinfo{volume}{29}},
  \bibinfo{pages}{376} (\bibinfo{year}{2004}).

\bibitem[{Joa(2005)}]{Joachim_Ratner05}
\bibinfo{journal}{Proc. Natl. Acad. Sci. USA. Special issue on molecular
  electronics, edited by C. Joachim and M.A.Ratner.}
  \textbf{\bibinfo{volume}{102}}, \bibinfo{pages}{8800} (\bibinfo{year}{2005}).

\bibitem[{\citenamefont{Kohler et~al.}(2002)\citenamefont{Kohler, Lehmann,
  Camalet, and Hanggi}}]{Hanggi02}
\bibinfo{author}{\bibfnamefont{S.}~\bibnamefont{Kohler}},
  \bibinfo{author}{\bibfnamefont{J.}~\bibnamefont{Lehmann}},
  \bibinfo{author}{\bibfnamefont{S.}~\bibnamefont{Camalet}}, \bibnamefont{and}
  \bibinfo{author}{\bibfnamefont{P.}~\bibnamefont{Hanggi}},
  \bibinfo{journal}{Israel J. of Chem.} \textbf{\bibinfo{volume}{42}},
  \bibinfo{pages}{135–} (\bibinfo{year}{2002}).

\bibitem[{\citenamefont{Kohler et~al.}(2004)\citenamefont{Kohler, Camalet,
  Strass, Lehmann, Ingold, and Hanggi}}]{Hanggi04}
\bibinfo{author}{\bibfnamefont{S.}~\bibnamefont{Kohler}},
  \bibinfo{author}{\bibfnamefont{S.}~\bibnamefont{Camalet}},
  \bibinfo{author}{\bibfnamefont{M.}~\bibnamefont{Strass}},
  \bibinfo{author}{\bibfnamefont{J.}~\bibnamefont{Lehmann}},
  \bibinfo{author}{\bibfnamefont{G.-L.} \bibnamefont{Ingold}},
  \bibnamefont{and} \bibinfo{author}{\bibfnamefont{P.}~\bibnamefont{Hanggi}},
  \bibinfo{journal}{Chem. Phys.} \textbf{\bibinfo{volume}{296}},
  \bibinfo{pages}{243–} (\bibinfo{year}{2004}).

\bibitem[{\citenamefont{Kim et~al.}(2007)\citenamefont{Kim, Kwon, and
  Kim}}]{Kim3}
\bibinfo{author}{\bibfnamefont{W.~Y.} \bibnamefont{Kim}},
  \bibinfo{author}{\bibfnamefont{S.~K.} \bibnamefont{Kwon}}, \bibnamefont{and}
  \bibinfo{author}{\bibfnamefont{K.~S.} \bibnamefont{Kim}},
  \bibinfo{journal}{Phys. Rev. B} \textbf{\bibinfo{volume}{76}},
  \bibinfo{pages}{033425} (\bibinfo{year}{2007}).

\bibitem[{\citenamefont{Diefenbach and Kim}(2007)}]{Dief}
\bibinfo{author}{\bibfnamefont{M.}~\bibnamefont{Diefenbach}} \bibnamefont{and}
  \bibinfo{author}{\bibfnamefont{K.~S.} \bibnamefont{Kim}},
  \bibinfo{journal}{Angew. Chem. Int.} \textbf{\bibinfo{volume}{Ed.46}},
  \bibinfo{pages}{7640} (\bibinfo{year}{2007}).

\bibitem[{\citenamefont{Dulic et~al.}(2003)\citenamefont{Dulic, van~der Molen,
  Kudernac, Jonkman, de~Jong, Bowden, van Esch, Feringa, and van
  Wees}}]{Dulic03}
\bibinfo{author}{\bibfnamefont{D.}~\bibnamefont{Dulic}},
  \bibinfo{author}{\bibfnamefont{S.}~\bibnamefont{van~der Molen}},
  \bibinfo{author}{\bibfnamefont{T.}~\bibnamefont{Kudernac}},
  \bibinfo{author}{\bibfnamefont{H.}~\bibnamefont{Jonkman}},
  \bibinfo{author}{\bibfnamefont{J.}~\bibnamefont{de~Jong}},
  \bibinfo{author}{\bibfnamefont{T.}~\bibnamefont{Bowden}},
  \bibinfo{author}{\bibfnamefont{J.}~\bibnamefont{van Esch}},
  \bibinfo{author}{\bibfnamefont{B.}~\bibnamefont{Feringa}}, \bibnamefont{and}
  \bibinfo{author}{\bibfnamefont{B.}~\bibnamefont{van Wees}},
  \bibinfo{journal}{Phys. Rev. Lett.} \textbf{\bibinfo{volume}{91}},
  \bibinfo{pages}{207402} (\bibinfo{year}{2003}).

\bibitem[{\citenamefont{Katsonis et~al.}(2006)\citenamefont{Katsonis, Kudernac,
  Walko, and et~al.}}]{Katsonis06}
\bibinfo{author}{\bibfnamefont{N.}~\bibnamefont{Katsonis}},
  \bibinfo{author}{\bibfnamefont{T.}~\bibnamefont{Kudernac}},
  \bibinfo{author}{\bibfnamefont{M.}~\bibnamefont{Walko}}, \bibnamefont{and}
  \bibinfo{author}{\bibnamefont{et~al.}}, \bibinfo{journal}{Adv. Materials}
  \textbf{\bibinfo{volume}{18}}, \bibinfo{pages}{1397} (\bibinfo{year}{2006}).

\bibitem[{\citenamefont{He et~al.}(2005)\citenamefont{He, Chen, Liddell, and
  et~al.}}]{He05}
\bibinfo{author}{\bibfnamefont{J.}~\bibnamefont{He}},
  \bibinfo{author}{\bibfnamefont{F.}~\bibnamefont{Chen}},
  \bibinfo{author}{\bibfnamefont{P.}~\bibnamefont{Liddell}}, \bibnamefont{and}
  \bibinfo{author}{\bibnamefont{et~al.}}, \bibinfo{journal}{Nanotechnology}
  \textbf{\bibinfo{volume}{16}}, \bibinfo{pages}{695} (\bibinfo{year}{2005}).

\bibitem[{\citenamefont{Wakayama et~al.}(2004)\citenamefont{Wakayama, Ogawa,
  Kubota, Suzuki, Kamikado, and Mashiko}}]{Wakayama04}
\bibinfo{author}{\bibfnamefont{Y.}~\bibnamefont{Wakayama}},
  \bibinfo{author}{\bibfnamefont{K.}~\bibnamefont{Ogawa}},
  \bibinfo{author}{\bibfnamefont{T.}~\bibnamefont{Kubota}},
  \bibinfo{author}{\bibfnamefont{H.}~\bibnamefont{Suzuki}},
  \bibinfo{author}{\bibfnamefont{T.}~\bibnamefont{Kamikado}}, \bibnamefont{and}
  \bibinfo{author}{\bibfnamefont{S.}~\bibnamefont{Mashiko}},
  \bibinfo{journal}{Appl. Phys. Lett.} \textbf{\bibinfo{volume}{85}},
  \bibinfo{pages}{329} (\bibinfo{year}{2004}).

\bibitem[{\citenamefont{Yasutomi et~al.}(2004)\citenamefont{Yasutomi, Morita,
  Imanishi, and Kimura}}]{Yasutomi04}
\bibinfo{author}{\bibfnamefont{S.}~\bibnamefont{Yasutomi}},
  \bibinfo{author}{\bibfnamefont{T.}~\bibnamefont{Morita}},
  \bibinfo{author}{\bibfnamefont{Y.}~\bibnamefont{Imanishi}}, \bibnamefont{and}
  \bibinfo{author}{\bibfnamefont{S.}~\bibnamefont{Kimura}},
  \bibinfo{journal}{Science} \textbf{\bibinfo{volume}{304}},
  \bibinfo{pages}{1944} (\bibinfo{year}{2004}).

\bibitem[{\citenamefont{Landauer}(1957)}]{Land1}
\bibinfo{author}{\bibfnamefont{R.}~\bibnamefont{Landauer}},
  \bibinfo{journal}{IBM Journal Res.Dev.} \textbf{\bibinfo{volume}{1}},
  \bibinfo{pages}{223} (\bibinfo{year}{1957}).

\bibitem[{\citenamefont{Landauer}(1970)}]{Landauer1}
\bibinfo{author}{\bibfnamefont{R.}~\bibnamefont{Landauer}},
  \bibinfo{journal}{Phil.Mag.} \textbf{\bibinfo{volume}{21}},
  \bibinfo{pages}{863} (\bibinfo{year}{1970}).

\bibitem[{\citenamefont{Kim and Kim}(2008{\natexlab{b}})}]{Kim2}
\bibinfo{author}{\bibfnamefont{W.~Y.} \bibnamefont{Kim}} \bibnamefont{and}
  \bibinfo{author}{\bibfnamefont{K.~S.} \bibnamefont{Kim}},
  \bibinfo{journal}{J. Comput. Chem.} \textbf{\bibinfo{volume}{29}},
  \bibinfo{pages}{1073} (\bibinfo{year}{2008}{\natexlab{b}}).

\bibitem[{\citenamefont{Kim et~al.}(2008)\citenamefont{Kim, Choi, and
  Kim}}]{Kim5}
\bibinfo{author}{\bibfnamefont{W.}~\bibnamefont{Kim}},
  \bibinfo{author}{\bibfnamefont{Y.}~\bibnamefont{Choi}}, \bibnamefont{and}
  \bibinfo{author}{\bibfnamefont{K.}~\bibnamefont{Kim}}, \bibinfo{journal}{J.
  Mater. Chem.} \textbf{\bibinfo{volume}{18}}, \bibinfo{pages}{4510}
  (\bibinfo{year}{2008}).

\bibitem[{\citenamefont{Cho et~al.}(2007)\citenamefont{Cho, Chen, and
  Fuhrerb}}]{Sungjae}
\bibinfo{author}{\bibfnamefont{S.}~\bibnamefont{Cho}},
  \bibinfo{author}{\bibfnamefont{Y.-F.} \bibnamefont{Chen}}, \bibnamefont{and}
  \bibinfo{author}{\bibfnamefont{M.~S.} \bibnamefont{Fuhrerb}},
  \bibinfo{journal}{Appl. Phys. Lett.} \textbf{\bibinfo{volume}{91}},
  \bibinfo{pages}{123105} (\bibinfo{year}{2007}).

\bibitem[{\citenamefont{Martins et~al.}(2007)\citenamefont{Martins, Miwa,
  da~Silva, and Fazzio}}]{Martins}
\bibinfo{author}{\bibfnamefont{T.~B.} \bibnamefont{Martins}},
  \bibinfo{author}{\bibfnamefont{R.~H.} \bibnamefont{Miwa}},
  \bibinfo{author}{\bibfnamefont{A.~J.~R.} \bibnamefont{da~Silva}},
  \bibnamefont{and} \bibinfo{author}{\bibfnamefont{A.}~\bibnamefont{Fazzio}},
  \bibinfo{journal}{Phys. Rev. Lett.} \textbf{\bibinfo{volume}{98}},
  \bibinfo{pages}{196803} (\bibinfo{year}{2007}).

\bibitem[{\citenamefont{Pisani et~al.}(2007)\citenamefont{Pisani, Chan,
  Montanari, and Harrison}}]{Pisani}
\bibinfo{author}{\bibfnamefont{L.}~\bibnamefont{Pisani}},
  \bibinfo{author}{\bibfnamefont{J.}~\bibnamefont{Chan}},
  \bibinfo{author}{\bibfnamefont{B.}~\bibnamefont{Montanari}},
  \bibnamefont{and} \bibinfo{author}{\bibfnamefont{N.}~\bibnamefont{Harrison}},
  \bibinfo{journal}{Phys.Rev.B} \textbf{\bibinfo{volume}{75}},
  \bibinfo{pages}{064418} (\bibinfo{year}{2007}).

\bibitem[{\citenamefont{Galperin and Nitzan}(2005)}]{Nit05}
\bibinfo{author}{\bibfnamefont{M.}~\bibnamefont{Galperin}} \bibnamefont{and}
  \bibinfo{author}{\bibfnamefont{A.}~\bibnamefont{Nitzan}},
  \bibinfo{journal}{Phys. Rev. Lett.} \textbf{\bibinfo{volume}{95}},
  \bibinfo{pages}{206802} (\bibinfo{year}{2005}).

\bibitem[{\citenamefont{M.Galperin and Nitzan}(2006)}]{Nitzan06JCP}
\bibinfo{author}{\bibnamefont{M.Galperin}} \bibnamefont{and}
  \bibinfo{author}{\bibfnamefont{A.}~\bibnamefont{Nitzan}},
  \bibinfo{journal}{J. Chem. Phys.} \textbf{\bibinfo{volume}{124}},
  \bibinfo{pages}{234709} (\bibinfo{year}{2006}).

\bibitem[{\citenamefont{Galperin et~al.}(2006)\citenamefont{Galperin, Nitzan,
  and Ratner}}]{Nit7}
\bibinfo{author}{\bibfnamefont{M.}~\bibnamefont{Galperin}},
  \bibinfo{author}{\bibfnamefont{A.}~\bibnamefont{Nitzan}}, \bibnamefont{and}
  \bibinfo{author}{\bibfnamefont{M.}~\bibnamefont{Ratner}},
  \bibinfo{journal}{Phys.Rev.B.} \textbf{\bibinfo{volume}{74}},
  \bibinfo{pages}{075326} (\bibinfo{year}{2006}).

\bibitem[{\citenamefont{B\"uttiker et~al.}(1985)\citenamefont{B\"uttiker, Imry,
  Landauer, and Pinhas}}]{Butt}
\bibinfo{author}{\bibfnamefont{M.}~\bibnamefont{B\"uttiker}},
  \bibinfo{author}{\bibfnamefont{Y.}~\bibnamefont{Imry}},
  \bibinfo{author}{\bibfnamefont{R.}~\bibnamefont{Landauer}}, \bibnamefont{and}
  \bibinfo{author}{\bibfnamefont{S.}~\bibnamefont{Pinhas}},
  \bibinfo{journal}{Phys. Rev. B} \textbf{\bibinfo{volume}{31}},
  \bibinfo{pages}{6207} (\bibinfo{year}{1985}).

\bibitem[{\citenamefont{B\"uttiker}(1986)}]{Butt2}
\bibinfo{author}{\bibfnamefont{M.}~\bibnamefont{B\"uttiker}},
  \bibinfo{journal}{Phys. Rev. Lett.} \textbf{\bibinfo{volume}{57}},
  \bibinfo{pages}{1761} (\bibinfo{year}{1986}).

\bibitem[{\citenamefont{Vager and Naaman}(2002)}]{Naa}
\bibinfo{author}{\bibfnamefont{Z.}~\bibnamefont{Vager}} \bibnamefont{and}
  \bibinfo{author}{\bibfnamefont{R.}~\bibnamefont{Naaman}},
  \bibinfo{journal}{Chem.Phys.} \textbf{\bibinfo{volume}{281}},
  \bibinfo{pages}{305} (\bibinfo{year}{2002}).

\bibitem[{\citenamefont{Brandbyge et~al.}(2002)\citenamefont{Brandbyge, Mozos,
  Ordejon, Taylor, and Stokbro}}]{Brand}
\bibinfo{author}{\bibfnamefont{M.}~\bibnamefont{Brandbyge}},
  \bibinfo{author}{\bibfnamefont{J.~L.} \bibnamefont{Mozos}},
  \bibinfo{author}{\bibfnamefont{P.}~\bibnamefont{Ordejon}},
  \bibinfo{author}{\bibfnamefont{J.}~\bibnamefont{Taylor}}, \bibnamefont{and}
  \bibinfo{author}{\bibfnamefont{K.}~\bibnamefont{Stokbro}},
  \bibinfo{journal}{Phys. Rev. B} \textbf{\bibinfo{volume}{65}},
  \bibinfo{pages}{165401} (\bibinfo{year}{2002}).

\bibitem[{\citenamefont{Larade et~al.}(2001)\citenamefont{Larade, Taylor,
  Mehrez, and Guo}}]{Larad}
\bibinfo{author}{\bibfnamefont{B.}~\bibnamefont{Larade}},
  \bibinfo{author}{\bibfnamefont{J.}~\bibnamefont{Taylor}},
  \bibinfo{author}{\bibfnamefont{H.}~\bibnamefont{Mehrez}}, \bibnamefont{and}
  \bibinfo{author}{\bibfnamefont{H.}~\bibnamefont{Guo}},
  \bibinfo{journal}{Phys.Rev.B} \textbf{\bibinfo{volume}{64}},
  \bibinfo{pages}{075420} (\bibinfo{year}{2001}).

\bibitem[{\citenamefont{Bixon and Jortner}(1999)}]{Jort}
\bibinfo{author}{\bibfnamefont{M.}~\bibnamefont{Bixon}} \bibnamefont{and}
  \bibinfo{author}{\bibfnamefont{J.}~\bibnamefont{Jortner}},
  \bibinfo{journal}{Adv. Chem. Phys.} \textbf{\bibinfo{volume}{106}},
  \bibinfo{pages}{35} (\bibinfo{year}{1999}).

\bibitem[{\citenamefont{Nitzan et~al.}(2002)\citenamefont{Nitzan, Galperin,
  Ingold, and Grabert}}]{NitGal}
\bibinfo{author}{\bibfnamefont{A.}~\bibnamefont{Nitzan}},
  \bibinfo{author}{\bibfnamefont{M.}~\bibnamefont{Galperin}},
  \bibinfo{author}{\bibfnamefont{G.}~\bibnamefont{Ingold}}, \bibnamefont{and}
  \bibinfo{author}{\bibfnamefont{H.}~\bibnamefont{Grabert}},
  \bibinfo{journal}{J.Chem.Phys.} \textbf{\bibinfo{volume}{117}},
  \bibinfo{pages}{10837} (\bibinfo{year}{2002}).

\bibitem[{\citenamefont{Nitzan et~al.}(2000)\citenamefont{Nitzan, Jortner,
  J.Wilkie, Burin, and Ratner}}]{NitJor}
\bibinfo{author}{\bibfnamefont{A.}~\bibnamefont{Nitzan}},
  \bibinfo{author}{\bibfnamefont{J.}~\bibnamefont{Jortner}},
  \bibinfo{author}{\bibnamefont{J.Wilkie}},
  \bibinfo{author}{\bibfnamefont{A.}~\bibnamefont{Burin}}, \bibnamefont{and}
  \bibinfo{author}{\bibfnamefont{M.}~\bibnamefont{Ratner}},
  \bibinfo{journal}{J.Phys.Chem.B} \textbf{\bibinfo{volume}{104}},
  \bibinfo{pages}{5661} (\bibinfo{year}{2000}).

\bibitem[{\citenamefont{Nitzan}(2001)}]{Nit10}
\bibinfo{author}{\bibfnamefont{A.}~\bibnamefont{Nitzan}},
  \bibinfo{journal}{Annu.Rev.Phys.Chem.} \textbf{\bibinfo{volume}{52}},
  \bibinfo{pages}{681} (\bibinfo{year}{2001}).

\bibitem[{\citenamefont{Caron et~al.}(1986)\citenamefont{Caron, G.Perluzzo,
  G.Bader, and Sanche}}]{Caron}
\bibinfo{author}{\bibfnamefont{L.}~\bibnamefont{Caron}},
  \bibinfo{author}{\bibnamefont{G.Perluzzo}},
  \bibinfo{author}{\bibnamefont{G.Bader}}, \bibnamefont{and}
  \bibinfo{author}{\bibfnamefont{L.}~\bibnamefont{Sanche}},
  \bibinfo{journal}{Phys.Rev.B} \textbf{\bibinfo{volume}{33}},
  \bibinfo{pages}{3027} (\bibinfo{year}{1986}).

\bibitem[{\citenamefont{Fainberg et~al.}(2007)\citenamefont{Fainberg,
  Jouravlev, and Nitzan}}]{Fain}
\bibinfo{author}{\bibfnamefont{B.~D.} \bibnamefont{Fainberg}},
  \bibinfo{author}{\bibfnamefont{M.}~\bibnamefont{Jouravlev}},
  \bibnamefont{and} \bibinfo{author}{\bibfnamefont{A.}~\bibnamefont{Nitzan}},
  \bibinfo{journal}{Phys.Rev.B} \textbf{\bibinfo{volume}{76}},
  \bibinfo{eid}{245329} (\bibinfo{year}{2007}).

\bibitem[{\citenamefont{Lehmann et~al.}(2003)\citenamefont{Lehmann, Camalet,
  Kohler, and Hanggi}}]{Hanggi03}
\bibinfo{author}{\bibfnamefont{J.}~\bibnamefont{Lehmann}},
  \bibinfo{author}{\bibfnamefont{S.}~\bibnamefont{Camalet}},
  \bibinfo{author}{\bibfnamefont{S.}~\bibnamefont{Kohler}}, \bibnamefont{and}
  \bibinfo{author}{\bibfnamefont{P.}~\bibnamefont{Hanggi}},
  \bibinfo{journal}{Chem. Phys. Lett.} \textbf{\bibinfo{volume}{368}},
  \bibinfo{pages}{282–} (\bibinfo{year}{2003}).

\bibitem[{\citenamefont{Keilmann and Hillenbrand}(2004)}]{Keilmann}
\bibinfo{author}{\bibfnamefont{F.}~\bibnamefont{Keilmann}} \bibnamefont{and}
  \bibinfo{author}{\bibfnamefont{R.}~\bibnamefont{Hillenbrand}},
  \bibinfo{journal}{Phil. Trans. R. Soc. Lond. A}
  \textbf{\bibinfo{volume}{362}}, \bibinfo{pages}{787} (\bibinfo{year}{2004}).

\bibitem[{\citenamefont{Wingreen and Meir}(1994)}]{MeirAnd}
\bibinfo{author}{\bibfnamefont{N.~S.} \bibnamefont{Wingreen}} \bibnamefont{and}
  \bibinfo{author}{\bibfnamefont{Y.}~\bibnamefont{Meir}},
  \bibinfo{journal}{Phys. Rev. B} \textbf{\bibinfo{volume}{49}},
  \bibinfo{pages}{11040} (\bibinfo{year}{1994}).

\bibitem[{\citenamefont{Anderson}(41)}]{Ander}
\bibinfo{author}{\bibfnamefont{P.}~\bibnamefont{Anderson}},
  \bibinfo{journal}{Phys. Rev.} \textbf{\bibinfo{volume}{124}},
  \bibinfo{pages}{6104} (\bibinfo{year}{41}).

\bibitem[{\citenamefont{G.D.Mahan}()}]{Mah}
\bibinfo{author}{\bibnamefont{G.D.Mahan}}, \eprint{\textit{Many-particle
  physics} (Plenum Press, New York, 1990)}.

\bibitem[{\citenamefont{L.P.Kadanoff}()}]{Kad}
\bibinfo{author}{\bibnamefont{L.P.Kadanoff}}, \eprint{\textit{Statistical
  physics. Statics, Dynamics and Renormalization} (Singapore,World Scientific,
  2000)}.

\bibitem[{\citenamefont{Kadanoff}(1969)}]{Kad2}
\bibinfo{author}{\bibfnamefont{L.}~\bibnamefont{Kadanoff}},
  \bibinfo{journal}{Phys.Rev.} \textbf{\bibinfo{volume}{188}},
  \bibinfo{pages}{859} (\bibinfo{year}{1969}).

\bibitem[{\citenamefont{G\"oppert et~al.}(2002)\citenamefont{G\"oppert,
  Galperin, Altshuler, and Grabert}}]{Gop}
\bibinfo{author}{\bibfnamefont{G.}~\bibnamefont{G\"oppert}},
  \bibinfo{author}{\bibfnamefont{Y.~M.} \bibnamefont{Galperin}},
  \bibinfo{author}{\bibfnamefont{B.~L.} \bibnamefont{Altshuler}},
  \bibnamefont{and} \bibinfo{author}{\bibfnamefont{H.}~\bibnamefont{Grabert}},
  \bibinfo{journal}{Phys. Rev. B} \textbf{\bibinfo{volume}{66}},
  \bibinfo{pages}{195328} (\bibinfo{year}{2002}).

\bibitem[{\citenamefont{Imamura et~al.}(2000)\citenamefont{Imamura, Kobayashi,
  Takahashi, and Maekawa}}]{Hirosha}
\bibinfo{author}{\bibfnamefont{H.}~\bibnamefont{Imamura}},
  \bibinfo{author}{\bibfnamefont{N.}~\bibnamefont{Kobayashi}},
  \bibinfo{author}{\bibfnamefont{S.}~\bibnamefont{Takahashi}},
  \bibnamefont{and} \bibinfo{author}{\bibfnamefont{S.}~\bibnamefont{Maekawa}},
  \bibinfo{journal}{Phys. Rev. Lett.} \textbf{\bibinfo{volume}{84}},
  \bibinfo{pages}{1003} (\bibinfo{year}{2000}).

\bibitem[{\citenamefont{Nitzan}()}]{Nit1}
\bibinfo{author}{\bibfnamefont{A.}~\bibnamefont{Nitzan}},
  \eprint{\textit{Chemical Dynamics in Condensed Phases. Relaxation, Transfer
  and Reaction in Condensed Molecular Systems} (Oxford University Press,
  2006)}.

\bibitem[{\citenamefont{Wolfle et~al.}(1970)\citenamefont{Wolfle, Brenig, and
  Gotze}}]{Wolf}
\bibinfo{author}{\bibfnamefont{P.}~\bibnamefont{Wolfle}},
  \bibinfo{author}{\bibfnamefont{W.}~\bibnamefont{Brenig}}, \bibnamefont{and}
  \bibinfo{author}{\bibfnamefont{W.}~\bibnamefont{Gotze}}, \bibinfo{journal}{Z.
  Phys.} \textbf{\bibinfo{volume}{235}}, \bibinfo{pages}{59}
  (\bibinfo{year}{1970}).

\bibitem[{\citenamefont{Grabert}()}]{Grab}
\bibinfo{author}{\bibfnamefont{H.}~\bibnamefont{Grabert}},
  \eprint{\textit{Projection Operator Techniques in Nonequilibrium Statistical
  Mechanics} (Springer, New York, 1982)}.

\bibitem[{\citenamefont{Berman and Mukamel}(2004)}]{Muk2}
\bibinfo{author}{\bibfnamefont{O.}~\bibnamefont{Berman}} \bibnamefont{and}
  \bibinfo{author}{\bibfnamefont{S.}~\bibnamefont{Mukamel}},
  \bibinfo{journal}{Phys.Phys.B.} \textbf{\bibinfo{volume}{69}},
  \bibinfo{pages}{155430} (\bibinfo{year}{2004}).

\bibitem[{\citenamefont{Welack et~al.}(2006)\citenamefont{Welack, Schreiber,
  and Kleinekathofer}}]{Schreiber06}
\bibinfo{author}{\bibfnamefont{S.}~\bibnamefont{Welack}},
  \bibinfo{author}{\bibfnamefont{M.}~\bibnamefont{Schreiber}},
  \bibnamefont{and}
  \bibinfo{author}{\bibfnamefont{U.}~\bibnamefont{Kleinekathofer}},
  \bibinfo{journal}{J. Chem. Phys.} \textbf{\bibinfo{volume}{124}},
  \bibinfo{pages}{044712} (\bibinfo{year}{2006}).

\bibitem[{\citenamefont{Lindberg and Koch}(1988)}]{Koch}
\bibinfo{author}{\bibfnamefont{M.}~\bibnamefont{Lindberg}} \bibnamefont{and}
  \bibinfo{author}{\bibfnamefont{S.~W.} \bibnamefont{Koch}},
  \bibinfo{journal}{Phys. Rev. B} \textbf{\bibinfo{volume}{38}},
  \bibinfo{pages}{3342} (\bibinfo{year}{1988}).

\bibitem[{\citenamefont{I.I.Rabi}(1937)}]{Rabi1}
\bibinfo{author}{\bibnamefont{I.I.Rabi}}, \bibinfo{journal}{Phys.Rev.}
  \textbf{\bibinfo{volume}{51}}, \bibinfo{pages}{652} (\bibinfo{year}{1937}).

\bibitem[{\citenamefont{Tran~Thoai and Haug}(1993)}]{Haug2}
\bibinfo{author}{\bibfnamefont{D.~B.} \bibnamefont{Tran~Thoai}}
  \bibnamefont{and} \bibinfo{author}{\bibfnamefont{H.}~\bibnamefont{Haug}},
  \bibinfo{journal}{Phys. Rev. B} \textbf{\bibinfo{volume}{47}},
  \bibinfo{pages}{3574} (\bibinfo{year}{1993}).

\bibitem[{\citenamefont{Schilp et~al.}(1994)\citenamefont{Schilp, Kuhn, and
  Mahler}}]{Shilp1}
\bibinfo{author}{\bibfnamefont{J.}~\bibnamefont{Schilp}},
  \bibinfo{author}{\bibfnamefont{T.}~\bibnamefont{Kuhn}}, \bibnamefont{and}
  \bibinfo{author}{\bibfnamefont{G.}~\bibnamefont{Mahler}},
  \bibinfo{journal}{Phys. Rev. B} \textbf{\bibinfo{volume}{50}},
  \bibinfo{pages}{5435} (\bibinfo{year}{1994}).

\bibitem[{\citenamefont{Schilp et~al.}(1995)\citenamefont{Schilp, Kuhn, and
  Mahler}}]{Shilp2}
\bibinfo{author}{\bibfnamefont{J.}~\bibnamefont{Schilp}},
  \bibinfo{author}{\bibfnamefont{T.}~\bibnamefont{Kuhn}}, \bibnamefont{and}
  \bibinfo{author}{\bibfnamefont{G.}~\bibnamefont{Mahler}},
  \bibinfo{journal}{Phys. Stat. Sol.} \textbf{\bibinfo{volume}{188}},
  \bibinfo{pages}{417} (\bibinfo{year}{1995}).

\bibitem[{\citenamefont{Butscher et~al.}(2005)\citenamefont{Butscher, Forstner,
  Waldmuller, and Knorr}}]{buts}
\bibinfo{author}{\bibfnamefont{S.}~\bibnamefont{Butscher}},
  \bibinfo{author}{\bibfnamefont{J.}~\bibnamefont{Forstner}},
  \bibinfo{author}{\bibfnamefont{I.}~\bibnamefont{Waldmuller}},
  \bibnamefont{and} \bibinfo{author}{\bibfnamefont{A.}~\bibnamefont{Knorr}},
  \bibinfo{journal}{Phys. Rev. B} \textbf{\bibinfo{volume}{72}},
  \bibinfo{eid}{045314} (\bibinfo{year}{2005}).

\bibitem[{\citenamefont{Allen and Eberly}()}]{All75}
\bibinfo{author}{\bibfnamefont{L.}~\bibnamefont{Allen}} \bibnamefont{and}
  \bibinfo{author}{\bibfnamefont{J.-H.} \bibnamefont{Eberly}},
  \eprint{\textit{Optical resonance and two-level atoms} (John Wiley \& Sons,
  New York, 1997)}.

\bibitem[{\citenamefont{Stievater et~al.}(2001)\citenamefont{Stievater, Li,
  Steel, Gammon, Katzer, Park, Piermarocchi, and Sham}}]{Stievater01}
\bibinfo{author}{\bibfnamefont{T.~H.} \bibnamefont{Stievater}},
  \bibinfo{author}{\bibfnamefont{X.}~\bibnamefont{Li}},
  \bibinfo{author}{\bibfnamefont{D.~G.} \bibnamefont{Steel}},
  \bibinfo{author}{\bibfnamefont{D.}~\bibnamefont{Gammon}},
  \bibinfo{author}{\bibfnamefont{D.~S.} \bibnamefont{Katzer}},
  \bibinfo{author}{\bibfnamefont{D.}~\bibnamefont{Park}},
  \bibinfo{author}{\bibfnamefont{C.}~\bibnamefont{Piermarocchi}},
  \bibnamefont{and} \bibinfo{author}{\bibfnamefont{L.~J.} \bibnamefont{Sham}},
  \bibinfo{journal}{Phys. Rev. Lett.} \textbf{\bibinfo{volume}{87}},
  \bibinfo{pages}{133603} (\bibinfo{year}{2001}).

\bibitem[{\citenamefont{Kamada et~al.}(2001)\citenamefont{Kamada, Gotoh,
  Temmyo, Takagahara, and Ando}}]{Kamada_Takagahara01}
\bibinfo{author}{\bibfnamefont{H.}~\bibnamefont{Kamada}},
  \bibinfo{author}{\bibfnamefont{H.}~\bibnamefont{Gotoh}},
  \bibinfo{author}{\bibfnamefont{J.}~\bibnamefont{Temmyo}},
  \bibinfo{author}{\bibfnamefont{T.}~\bibnamefont{Takagahara}},
  \bibnamefont{and} \bibinfo{author}{\bibfnamefont{H.}~\bibnamefont{Ando}},
  \bibinfo{journal}{Phys. Rev. Lett.} \textbf{\bibinfo{volume}{87}},
  \bibinfo{pages}{246401} (\bibinfo{year}{2001}).

\bibitem[{\citenamefont{Htoon et~al.}(2001)\citenamefont{Htoon, Takagahara,
  Kulik, Baklenov, A.~L.~Holmes, and Shih}}]{Htoon_Takagahara02}
\bibinfo{author}{\bibfnamefont{H.}~\bibnamefont{Htoon}},
  \bibinfo{author}{\bibfnamefont{T.}~\bibnamefont{Takagahara}},
  \bibinfo{author}{\bibfnamefont{D.}~\bibnamefont{Kulik}},
  \bibinfo{author}{\bibfnamefont{O.}~\bibnamefont{Baklenov}},
  \bibinfo{author}{\bibfnamefont{J.}~\bibnamefont{A.~L.~Holmes}},
  \bibnamefont{and} \bibinfo{author}{\bibfnamefont{C.~K.} \bibnamefont{Shih}},
  \bibinfo{journal}{Phys. Rev. Lett.} \textbf{\bibinfo{volume}{88}},
  \bibinfo{pages}{087401} (\bibinfo{year}{2001}).

\bibitem[{\citenamefont{Zrenner et~al.}(2002)\citenamefont{Zrenner, Beham,
  Stufler, Findeis, Bichler, and Abstreiter}}]{Zrenner02Nature}
\bibinfo{author}{\bibfnamefont{A.}~\bibnamefont{Zrenner}},
  \bibinfo{author}{\bibfnamefont{E.}~\bibnamefont{Beham}},
  \bibinfo{author}{\bibfnamefont{S.}~\bibnamefont{Stufler}},
  \bibinfo{author}{\bibfnamefont{F.}~\bibnamefont{Findeis}},
  \bibinfo{author}{\bibfnamefont{M.}~\bibnamefont{Bichler}}, \bibnamefont{and}
  \bibinfo{author}{\bibfnamefont{G.}~\bibnamefont{Abstreiter}},
  \bibinfo{journal}{Nature} \textbf{\bibinfo{volume}{418}},
  \bibinfo{pages}{612} (\bibinfo{year}{2002}).

\bibitem[{\citenamefont{Shore et~al.}(1992)\citenamefont{Shore, Bergmann, Kuhn,
  Schiemann, Oreg, and Eberly}}]{Sho92}
\bibinfo{author}{\bibfnamefont{B.~W.} \bibnamefont{Shore}},
  \bibinfo{author}{\bibfnamefont{K.}~\bibnamefont{Bergmann}},
  \bibinfo{author}{\bibfnamefont{A.}~\bibnamefont{Kuhn}},
  \bibinfo{author}{\bibfnamefont{S.}~\bibnamefont{Schiemann}},
  \bibinfo{author}{\bibfnamefont{J.}~\bibnamefont{Oreg}}, \bibnamefont{and}
  \bibinfo{author}{\bibfnamefont{J.~H.} \bibnamefont{Eberly}},
  \bibinfo{journal}{Phys. Rev. A} \textbf{\bibinfo{volume}{45}},
  \bibinfo{pages}{5297} (\bibinfo{year}{1992}).

\bibitem[{\citenamefont{Melinger et~al.}(1994)\citenamefont{Melinger, Gandhi,
  Hariharan, Goswami, and Warren}}]{Mel94}
\bibinfo{author}{\bibfnamefont{J.~S.} \bibnamefont{Melinger}},
  \bibinfo{author}{\bibfnamefont{S.~R.} \bibnamefont{Gandhi}},
  \bibinfo{author}{\bibfnamefont{A.}~\bibnamefont{Hariharan}},
  \bibinfo{author}{\bibfnamefont{D.}~\bibnamefont{Goswami}}, \bibnamefont{and}
  \bibinfo{author}{\bibfnamefont{W.~S.} \bibnamefont{Warren}},
  \bibinfo{journal}{J. Chem. Phys.} \textbf{\bibinfo{volume}{101}},
  \bibinfo{pages}{6439} (\bibinfo{year}{1994}).

\bibitem[{\citenamefont{Treacy}(1968)}]{Tre68}
\bibinfo{author}{\bibfnamefont{E.~B.} \bibnamefont{Treacy}},
  \bibinfo{journal}{Phys. Lett. A} \textbf{\bibinfo{volume}{27}},
  \bibinfo{pages}{421} (\bibinfo{year}{1968}).

\bibitem[{\citenamefont{Vitanov et~al.}(2001)\citenamefont{Vitanov, Halfmann,
  Shore, and Bergmann}}]{Vit01}
\bibinfo{author}{\bibfnamefont{N.~V.} \bibnamefont{Vitanov}},
  \bibinfo{author}{\bibfnamefont{T.}~\bibnamefont{Halfmann}},
  \bibinfo{author}{\bibfnamefont{B.~W.} \bibnamefont{Shore}}, \bibnamefont{and}
  \bibinfo{author}{\bibfnamefont{K.}~\bibnamefont{Bergmann}},
  \bibinfo{journal}{Annu. Rev. Phys. Chem.} \textbf{\bibinfo{volume}{52}},
  \bibinfo{pages}{763} (\bibinfo{year}{2001}).

\bibitem[{\citenamefont{Fainberg and Gorbunov}(2004)}]{Fai04JCP}
\bibinfo{author}{\bibfnamefont{B.~D.} \bibnamefont{Fainberg}} \bibnamefont{and}
  \bibinfo{author}{\bibfnamefont{V.~A.} \bibnamefont{Gorbunov}},
  \bibinfo{journal}{J. Chem. Phys.} \textbf{\bibinfo{volume}{121}},
  \bibinfo{pages}{8748} (\bibinfo{year}{2004}).

\bibitem[{\citenamefont{Fainberg et~al.}(2005)\citenamefont{Fainberg, Levinsky,
  and Gorbunov}}]{Fai05JOSAB}
\bibinfo{author}{\bibfnamefont{B.~D.} \bibnamefont{Fainberg}},
  \bibinfo{author}{\bibfnamefont{B.}~\bibnamefont{Levinsky}}, \bibnamefont{and}
  \bibinfo{author}{\bibfnamefont{V.~A.} \bibnamefont{Gorbunov}},
  \bibinfo{journal}{J. Opt. Soc. Am. B} \textbf{\bibinfo{volume}{22}},
  \bibinfo{pages}{2715} (\bibinfo{year}{2005}).

\bibitem[{\citenamefont{Fainberg}(1998)}]{Fai98}
\bibinfo{author}{\bibfnamefont{B.~D.} \bibnamefont{Fainberg}},
  \bibinfo{journal}{J. Chem. Phys.} \textbf{\bibinfo{volume}{109}},
  \bibinfo{pages}{4523} (\bibinfo{year}{1998}).

\bibitem[{\citenamefont{Wang and Shen}(2006)}]{Wang_Shen06}
\bibinfo{author}{\bibfnamefont{F.}~\bibnamefont{Wang}} \bibnamefont{and}
  \bibinfo{author}{\bibfnamefont{Y.~R.} \bibnamefont{Shen}},
  \bibinfo{journal}{Phys. Rev. Lett.} \textbf{\bibinfo{volume}{97}},
  \bibinfo{pages}{206806} (\bibinfo{year}{2006}).

\bibitem[{\citenamefont{Brixner et~al.}(2006)\citenamefont{Brixner, de~Abajo,
  Schneider, Spindler, and Pfeiffer}}]{Brixner06}
\bibinfo{author}{\bibfnamefont{T.}~\bibnamefont{Brixner}},
  \bibinfo{author}{\bibfnamefont{F.~J.~G.} \bibnamefont{de~Abajo}},
  \bibinfo{author}{\bibfnamefont{J.}~\bibnamefont{Schneider}},
  \bibinfo{author}{\bibfnamefont{C.}~\bibnamefont{Spindler}}, \bibnamefont{and}
  \bibinfo{author}{\bibfnamefont{W.}~\bibnamefont{Pfeiffer}},
  \bibinfo{journal}{Phys. Rev. B} \textbf{\bibinfo{volume}{73}},
  \bibinfo{pages}{125437} (\bibinfo{year}{2006}).

\bibitem[{\citenamefont{Akulin and Schleich}(1992)}]{Aku92}
\bibinfo{author}{\bibfnamefont{V.~M.} \bibnamefont{Akulin}} \bibnamefont{and}
  \bibinfo{author}{\bibfnamefont{W.~P.} \bibnamefont{Schleich}},
  \bibinfo{journal}{Phys. Rev. A} \textbf{\bibinfo{volume}{46}},
  \bibinfo{pages}{4110} (\bibinfo{year}{1992}).

\bibitem[{\citenamefont{Petek and Ogawa}(1997)}]{Petek97}
\bibinfo{author}{\bibfnamefont{H.}~\bibnamefont{Petek}} \bibnamefont{and}
  \bibinfo{author}{\bibfnamefont{S.}~\bibnamefont{Ogawa}},
  \bibinfo{journal}{Progress in Surface Science} \textbf{\bibinfo{volume}{56}},
  \bibinfo{pages}{239} (\bibinfo{year}{1997}).

\bibitem[{\citenamefont{Galperin et~al.}(2007)\citenamefont{Galperin, Ratner,
  and Nitzan}}]{Nitzan07Jphys}
\bibinfo{author}{\bibfnamefont{M.}~\bibnamefont{Galperin}},
  \bibinfo{author}{\bibfnamefont{M.~A.} \bibnamefont{Ratner}},
  \bibnamefont{and} \bibinfo{author}{\bibfnamefont{A.}~\bibnamefont{Nitzan}},
  \bibinfo{journal}{J. Phys.: Condens. Matter} \textbf{\bibinfo{volume}{19}},
  \bibinfo{pages}{103201} (\bibinfo{year}{2007}).

\end{thebibliography}
\end{document}